\documentclass[a4paper,11pt]{article}
\usepackage[compatibility=false]{caption}
\usepackage{jheppub} 
\usepackage{lineno}
\usepackage{braket}
\usepackage{cancel}
\usepackage{mathrsfs}
\usepackage{tikz}
\usepackage{tikz-feynman}
\usetikzlibrary{decorations.pathmorphing} 
\usepackage{bm}
\usepackage{braket}
\usepackage{comment}
\usepackage{amsmath, amssymb}
\usepackage[compat=1.1.0]{tikz-feynhand}
\usepackage[subrefformat=parens]{subcaption}
\usepackage{ulem} 
\usepackage{physics}

\usepackage{cancel}

\newcommand{\CPV}{\cancel{\mathrm{CP}}}

\title{Systematic derivation of the Boltzmann equation for electroweak baryogenesis}

\makeatletter
\gdef\@fpheader{}
\makeatother

\preprint{KEK-TH-2855}

\author[a,b]{Motoi Endo,}

\author[a]{Yushi Mura,}

\author[a,b]{Tenta Tsuji}

\affiliation[a]{KEK Theory Center, Tsukuba, Ibaraki 305--0801, Japan}
\affiliation[b]{Graduate Institute for Advanced Studies, SOKENDAI, Tsukuba, Ibaraki 305--0801, Japan}

\abstract{
We derive the Boltzmann equation for electroweak baryogenesis within a field-theoretical framework.
We first algebraically solve the Kadanoff--Baym equations by incorporating the effects of gradients of the CP-violating bubble-wall background order by order.
We then apply approximations and assumptions appropriate for electroweak baryogenesis and identify the on-shell parts of the solutions, which describe quasiparticle states propagating in a plasma.
Using these on-shell solutions, we derive the Boltzmann equations for flavor-diagonal quasiparticle modes, while neglecting the off-diagonal components describing quantum coherence between different flavor modes.
Our derivation generalizes previous field-theoretical treatments by retaining self-energy corrections, and our results are consistent with the known Boltzmann equations when these corrections are neglected.
This framework provides a systematic and transparent derivation applicable to multi-flavor bosonic and fermionic systems.
}

\begin{document}
\maketitle
\flushbottom

\allowdisplaybreaks

\section{Introduction \label{sec:intro}}

Electroweak baryogenesis (EWBG)~\cite{Kuzmin:1985mm} is an attractive mechanism for explaining the baryon asymmetry of the Universe.
In this scenario, the electroweak phase transition is first order, and bubbles of the broken phase nucleate and expand in the surrounding symmetric phase.
The departure from equilibrium required for baryogenesis is realized around the expanding bubble wall.
If interactions between particles and the wall violate the CP symmetry, particles and antiparticles propagate differently in the wall, and charge asymmetries are generated.
The asymmetries present in the symmetric phase are then partially converted into a baryon asymmetry by electroweak sphaleron transitions.

A quantitative prediction of EWBG requires describing how the charge asymmetries generated near the wall are transported through the plasma and modified by scatterings and reactions.
In conventional analyses, particles passing through the wall are treated as quasiparticles moving in a slowly varying wall-background, and the evolution of their distribution functions is described by the Boltzmann equation~\cite{Cohen:1994ss,Joyce:1994fu,Joyce:1994zt,Huet:1995sh}.
In detail, the vacuum expectation values (VEVs) of scalar fields vary across the wall, so that particle masses become spacetime dependent and the quasiparticles experience a force from the wall.
If the mass term contains a CP-violating phase, this force differs between particles and antiparticles and acts as a source of the charge asymmetries.
This treatment is commonly referred to as the semiclassical force approach, and the resulting Boltzmann equation has provided the standard framework for describing transport phenomena in EWBG~\cite{Cline:1997vk,Cline:2000nw,Kainulainen:2001cn,Kainulainen:2002th,Prokopec:2003pj,Prokopec:2004ic,Fromme:2006wx,Cline:2020jre,Kainulainen:2021oqs,Kainulainen:2024qpm}.

The semiclassical Boltzmann description can be derived from a field-theoretical framework.
The statistical state of the non-equilibrium system is described in terms of correlation functions in the Schwinger--Keldysh formalism~\cite{Schwinger:1960qe,Keldysh:1964ud}.
In particular, the Wightman functions contain both the spectral and statistical information of the plasma.
Their dynamics is governed by the Kadanoff--Baym equations~\cite{Kadanoff:1962}.
Starting from these equations, one can derive the semiclassical force and the corresponding Boltzmann equation used in EWBG calculations~\cite{Kainulainen:2001cn,Kainulainen:2002th,Prokopec:2003pj}.

The Kadanoff--Baym equations are separated into constraint and kinetic equations~\cite{Kainulainen:2001cn,Kainulainen:2002th,Prokopec:2003pj}.
The constraint equation determines the dispersion relation of the quasiparticle, or equivalently the on-shell conditions for the Wightman functions.
On the other hand, the kinetic equation describes the evolution of the corresponding distribution functions, and it leads to the semiclassical Boltzmann equation after substituting the on-shell solution.
When the bubble wall varies slowly compared with the temperature of the plasma, the effect of the wall can be incorporated through a derivative expansion.
For fermions with a complex spacetime-dependent mass, the gradient of the mass phase induces a CP-odd correction to the quasiparticle dispersion relation at first order in the derivative expansion.
This correction appears in the Boltzmann equation as a CP-violating semiclassical force.
For bosons, on the other hand, no analogous CP-odd correction to the dispersion relation appears at this order, and hence no CP-violating force term is generated in the conventional semiclassical analysis~\cite{Kainulainen:2001cn,Prokopec:2003pj}.
These results are also consistent with the quantum-mechanical derivation based on the WKB approximation~\cite{Cline:1997vk,Cline:2000nw,Fromme:2006wx}.
Moreover, the Kadanoff--Baym framework can also describe flavor mixing and quantum coherence between different flavor states~\cite{Konstandin:2004gy,Konstandin:2005cd,Herranen:2008hi,Cirigliano:2009yt,Cirigliano:2011di}.\footnote{
Another framework is the so-called VEV-insertion-approximation (VIA) method~\cite{Riotto:1995hh,Riotto:1997vy,Riotto:1998zb}.
In appendix~\ref{sec:VIA}, we briefly review the VIA method and recent criticisms of it~\cite{Kainulainen:2021oqs,Postma:2022dbr}, and discuss how these issues are understood in our formulation.
}

In the conventional derivations~\cite{Kainulainen:2001cn,Kainulainen:2002th,Prokopec:2003pj}, however, self-energy corrections are not retained when solving the Kadanoff--Baym equations.
The corresponding on-shell solution therefore essentially describes particles propagating through a spacetime-dependent background in vacuum, possibly including corrections induced by the wall gradients.
However, it does not fully describe the propagation of thermally dressed quasiparticles in the plasma.
To account for these thermal effects, the self-energy corrections that determine their dispersion relations and damping properties should be incorporated consistently into the derivation of the Boltzmann equation.

Motivated by this issue, in the present work we derive the Boltzmann equation including self-energy corrections for both bosonic and fermionic systems, thereby generalizing previous derivations~\cite{Kainulainen:2001cn,Kainulainen:2002th,Prokopec:2003pj}.
We start by algebraically solving the Kadanoff--Baym equations with self-energy corrections.
We then identify the on-shell parts of the solutions by applying approximations and assumptions appropriate for the EWBG scenario.
Projecting these on-shell solutions onto the kinetic equation leads to the Boltzmann equations for the corresponding distribution functions.
In this work, we focus on flavor-diagonal quasiparticle modes and do not retain the off-diagonal components describing quantum coherence between different flavor modes.
We show that, for fermions, the correction to the quasiparticle shell at first order in the derivative expansion leads to CP-violating particle-antiparticle differences in the group velocity and semiclassical force terms of the Boltzmann equation.
For bosons, no analogous CP-odd correction to the quasiparticle shell appears at this order, and hence no corresponding CP-violating terms are generated in the semiclassical Boltzmann equation.

We now comment on the relation between our approach and the recent work in ref.~\cite{Kainulainen:2021oqs}, in which thermal corrections, including finite-width effects, were discussed in a one-flavor fermionic system.
In that analysis, a specific form of the self-energy was employed, and the finite-width effects were treated within a thermal WKB quasiparticle picture through the modification of the pole structure entering the semiclassical source.
The main difference from the present work lies in the derivation of the Boltzmann equation: our formulation starts directly from the analytic solutions of the Kadanoff--Baym equations with self-energy corrections.
This formulation can be applied systematically to multi-flavor bosonic and fermionic systems without specifying an explicit model-dependent expression for the self-energy.
It also provides a unified perspective on the thermally corrected quasiparticle dispersion relations, the statistical information carried by the Wightman functions, and the collision terms.
In addition, it makes explicit the assumptions required to derive the usual semiclassical Boltzmann equation.

The organization of this paper is as follows.
In section~\ref{sec:KB}, we introduce the Kadanoff--Baym equations in the Schwinger--Keldysh formalism and define the propagators and self-energy used throughout the paper.
We also introduce the local mass eigenbasis that diagonalizes the spacetime-dependent mass matrix in flavor space.
In section~\ref{sec:approx}, we clarify the approximations and assumptions relevant for EWBG, including the derivative expansion, the planar-wall approximation, the steady-wall assumption, the quasiparticle approximation, the perturbation for flavor mixing, and the decomposition of the Wightman self-energy.
In section~\ref{sec:solution}, we solve the Kadanoff--Baym equations algebraically in the derivative expansion.
In section~\ref{sec:systematic}, we identify the on-shell part of the spectral and Wightman functions and derive the Boltzmann equation for bosons and fermions.
We conclude in section~\ref{sec:conclusions}.
Several technical details and comments on the VIA method are collected in the appendices.

\section{Kadanoff--Baym equations \label{sec:KB}}

In this section, we derive the Kadanoff--Baym equations for a non-equilibrium quantum field theory in the Schwinger--Keldysh formalism.
In this formalism, operators are defined on the closed-time path (CTP) $C = C_+ \cup C_-$.
The branch $C_+$ runs from past infinity to a finite time $t$, while the branch $C_-$ returns from $t$ to past infinity.

\subsection{Propagators}
On the CTP, the contour-ordered propagators for complex scalar bosons $\phi$ and four-component Dirac fermions $\psi$ are defined by 
\begin{align}
  i \Delta_{ij} (u,v) &\equiv \langle \mathrm{T}_C [\phi_i(u) \phi_j^\dagger(v)] \rangle, \notag \\
  i S_{ij,\alpha \beta} (u,v) &\equiv \langle \mathrm{T}_C [\psi_{i, \alpha}(u) \overline{\psi}_{j, \beta}(v)] \rangle,
\end{align} 
respectively, where $u,v$ are the spacetime variables, $i,j$ are the flavor indices, and $\alpha, \beta$ are the spinor indices.
There are four two-point Green functions: for the complex scalar bosons, they are given by
\begin{align}
    &i\Delta ^{++}_{ij} (u,v) \equiv i\Delta^{t}_{ij} (u,v) = \langle \mathrm{T} [\phi_i (u) \phi^\dagger_j (v)] \rangle, \notag \\
    &i \Delta ^{+-}_{ij} (u,v) \equiv  i\Delta^{<}_{ij} (u,v) =  \langle \phi^\dagger_j (v) \phi_i (u) \rangle, \notag \\
    &i\Delta ^{-+}_{ij} (u,v) \equiv  i\Delta^{>}_{ij} (u,v) =  \langle \phi_i (u) \phi^\dagger_j (v) \rangle, \notag \\
    &i\Delta ^{--}_{ij} (u,v) \equiv  i\Delta^{\overline{t}}_{ij} (u,v) = \langle \mathrm{\overline{T}} [\phi_i (u) \phi^\dagger_j (v)] \rangle.
    \label{eq:defboson}
\end{align}
For the fermions, we have
\begin{align} 
    &iS_{ij,\alpha \beta} ^{++} (u,v) \equiv iS_{ij,\alpha \beta} ^{t} (u,v) =  \langle \mathrm{T} [\psi_{i, \alpha} (u) \overline{\psi}_{j, \beta} (v)] \rangle, \notag \\
    &iS_{ij,\alpha \beta} ^{+-} (u,v) \equiv  iS_{ij,\alpha \beta} ^{<} (u,v) = - \langle \overline{\psi}_{j, \beta} (v) \psi_{i, \alpha} (u) \rangle, \notag \\
    &iS_{ij,\alpha \beta} ^{-+} (u,v) \equiv  iS_{ij,\alpha \beta} ^{>} (u,v) =  \langle \psi_{i, \alpha} (u) \overline{\psi}_{j, \beta} (v) \rangle, \notag \\
    &iS_{ij,\alpha \beta} ^{--} (u,v) \equiv  iS_{ij,\alpha \beta} ^{\overline{t}} (u,v) =  \langle \mathrm{\overline{T}} [\psi_{i, \alpha} (u) \overline{\psi}_{j, \beta} (v)] \rangle.
    \label{eq:deffermion}
\end{align}
The anti-time ordering product is denoted by $\overline{\mathrm{T}}[...]$.
In the following, we use the collective notation $G_{ij}^{X}=\Delta_{ij}^{X}$ for bosons and $G_{ij,\alpha\beta}^{X} = S_{ij,\alpha\beta}^{X}$ for fermions, where $X \in \{<,>,t,\overline{t} \}$.
We often suppress the flavor and spinor indices unless they are needed explicitly.
The propagators $G^<$ and $G^>$ are referred to as the Wightman functions.
The time-ordered and anti-time-ordered propagators, $G^t$ and $G^{\overline{t}}$, are expressed in terms of the Wightman functions as
\begin{align}
  &G^{t}(u,v) = \theta(u^0 - v^0) G^> (u,v) + \theta(v^0 - u^0) G^< (u,v), \notag \\
  &G^{\overline{t}}(u,v) = \theta(u^0 - v^0) G^< (u,v) + \theta(v^0 - u^0) G^> (u,v).
  \label{eq:propag1}
\end{align}
The retarded and advanced propagators are defined by
\begin{align}
    G^r \equiv G^t - G^< = G^> - G^{\overline{t}}, \notag \\
    G^a \equiv G^t - G^> = G^< - G^{\overline{t}},
    \label{eq:propag2}
\end{align}
and their Hermitian and anti-Hermitian parts are defined as
\begin{align}
  &G_{\mathcal{H}} \equiv \frac{1}{2} (G^r + G^a), \notag \\
  &G_{\mathcal{A}} \equiv \frac{1}{2i} (G^a - G^r) = \frac{i}{2} (G^> - G^<) \equiv \mathscr{A}.
  \label{eq:propag3}
\end{align}
The anti-Hermitian part $\mathscr{A}$ describes the spectral density of the plasma and is therefore called the spectral function.

\subsection{Kadanoff--Baym equations}

The Schwinger--Dyson equation on the CTP is given by~\cite{Prokopec:2003pj} 
\begin{align}
  K(u) i G^{ab}(u,v) = ia \delta^{ab} \delta^4(u-v) + \sum_c c \int d^4w~ g^{ac}(u,w) iG^{cb}(w, v),
  \label{eq:SD1}
\end{align}
where $a,b,c = \pm$, and 
\begin{equation}
    K(u) \equiv 
    \begin{cases}
     -\big( \Box_u + M^2(u) \big)~&(\mathrm{bosons}) \\
     i\cancel{\partial}_u - M_0(u) - i \gamma^5 M_5(u) ~&(\mathrm{fermions})
    \end{cases}.
\end{equation}
The spacetime-dependent masses $M^2(u)$, $M_0(u)$, and $M_5 (u)$ are, in general, Hermitian matrices in flavor space, while the kinetic operators are taken to be flavor diagonal.
In section~\ref{sec:basis}, we discuss in more detail the basis transformation that diagonalizes the mass matrix.

We use the collective notation $g^{ab}$ for the self-energy, with $g^{ab}=\Pi^{ab}$ for bosons and $g^{ab}=\Sigma^{ab}$ for fermions.
The self-energy components are labeled in the same way as the propagators; for example, $g^{++}=g^t$, and relations analogous to Eqs.~\eqref{eq:propag1}-\eqref{eq:propag3} hold for the self-energy. 
The Hermitian and anti-Hermitian parts of the self-energy are denoted by $g_{\mathcal H}$ and $g_{\mathcal A}$, respectively.
Physically, $g_{\mathcal H}$ corresponds to the dispersive part and gives a correction to the quasiparticle dispersion relation, while $g_{\mathcal A}$ corresponds to the absorptive part.
Using these relations, Eq.~\eqref{eq:SD1} yields
\begin{align}
  K(u) G^{r,a} (u,v) - [g^{r,a} \odot G^{r,a}](u,v) &= \delta^4 (u-v), \label{eq:SDequation1} \\
  K(u) G^{\lambda} (u,v) - [g_{\mathcal{H}} \odot G^{\lambda}](u,v) - [g^{\lambda} \odot G_{\mathcal{H}}](u,v)  &= \frac{1}{2}\Big( [g^> \odot G^<] - [g^< \odot G^>] \Big)(u,v),
  \label{eq:SDequation2}
\end{align}
where $\lambda = <,>$, and
\begin{align}
    [A \odot B] (u,v) \equiv \int d^4w~ A(u,w) B(w, v).
\end{align}

To describe a non-equilibrium system, it is useful to introduce the relative and average coordinates, $r = u-v$ and $x = (u+v)/2$.
These coordinates describe microscopic plasma dynamics and macroscopic variations, respectively.
The Wigner transformation for a function $f(u,v)$ is defined by 
\begin{align}
  f(k,x) = \int d^4 r ~ e^{ikr} f(u = x + r/2, v = x - r/2).
  \label{eq:Wigner1}
\end{align}
Applying the Wigner transformation to Eqs.~\eqref{eq:SDequation1}-\eqref{eq:SDequation2}, we obtain the Kadanoff--Baym equations
\begin{align}
  \mathcal{D}(k,x) G^{r,a} - e^{-i\diamond}\big( \mathcal{M} + g^{r,a} ,   G^{r,a} \big) &= 1, \label{eq:propag} \\
  \mathcal{D}(k,x) G^{\lambda} - e^{-i\diamond}\big( \mathcal{M} + g_{\mathcal{H}} , G^{\lambda} \big) - e^{-i\diamond} \big( g^{\lambda},  G_{\mathcal{H}} \big) &= \frac{1}{2} \Big( e^{-i\diamond} \big( g^>,  G^< \big) - e^{-i\diamond} \big( g^< , G^> \big) \Big).
  \label{eq:Wightman1}
\end{align}
The diamond operator is defined by 
\begin{align}
    e^{- i \diamond} \big(A(k,x) ,B(k,x) \big) \equiv \sum_{n=0}^\infty \frac{(-i)^n}{n !} \diamond^n \big(A(k,x) ,B(k,x) \big),
    \label{eq:exp_expand}
\end{align}
where 
\begin{align}
  \diamond^n \big(A(k,x) ,B(k,x) \big)
  \equiv
  \frac{1}{2^n} \big( \partial_x^A\cdot\partial_k^B - \partial_k^A\cdot\partial_x^B \big)^n  A(k,x) B(k,x) .
\end{align}
The Lorentz indices are suppressed here.
For both bosonic and fermionic cases, $\mathcal{D}$ and $\mathcal{M}$ are written by
\begin{align}
    \mathcal{D}(k,x) \equiv 
    \begin{cases}
     k^2 - \frac{1}{4} \partial_x^2 + ik \cdot \partial_x \\
     \cancel{k} + \frac{i}{2} \cancel{\partial}_x
    \end{cases}, ~~
    \mathcal{M} \equiv
    \begin{cases}
     M^2(x) ~~~&(\mathrm{bosons}) \\
    M_0(x) + i \gamma^5 M_5(x) ~~~&(\mathrm{fermions})
    \end{cases}.
\end{align}
Notably, Eq.~\eqref{eq:Wightman1} can be equivalently rewritten in the simpler form
\begin{align}
    \mathcal{D}(k,x) G^\lambda - e^{-i \diamond} \Big( \big( \mathcal{M} + g_{\mathcal{H}} - ig_{\mathcal{A}} \big), G^\lambda \Big) - e^{-i \diamond} \big( g^\lambda, G^a  \big) = 0.
    \label{eq:Wightman2}
\end{align}

The equations for the Wightman functions in Eq.~\eqref{eq:Wightman1} are further divided into the two equations, the Hermitian and anti-Hermitian parts.
By taking the Hermitian conjugate\footnote{
    For fermions, the corresponding conjugate is given by the Dirac conjugate, $A^{(\mathrm{d.c.})} = \gamma^0 A^\dagger \gamma^0$.
} of Eq.~\eqref{eq:Wightman1}, we have the anti-Hermitian and Hermitian parts, 
\begin{align}
  &\{\mathcal{D}_C, G^\lambda \} +[\mathcal{D}_K, G^\lambda] - e^{-i \diamond} \Big( \{ \mathcal{M} + g_{\mathcal{H}} ,G^\lambda \} +\{ g^{\lambda}, G_{\mathcal{H}} \} \Big) = \frac{1}{2} e^{-i\diamond} \Big( [g^>, G^<] - [g^<, G^>] \Big), \label{eq:KBeq_0} \\
  &[\mathcal{D}_C, G^\lambda ] +\{\mathcal{D}_K, G^\lambda \} - e^{-i \diamond} \Big( [ \mathcal{M} + g_{\mathcal{H}} ,G^\lambda ] +[ g^{\lambda}, G_{\mathcal{H}} ] \Big) = \frac{1}{2} e^{-i\diamond} \Big( \{g^>, G^<\} - \{g^<, G^>\} \Big),
  \label{eq:KBeq_1}
\end{align}
respectively.
We have introduced the commutator and anticommutator,
$[A,B] = AB - BA$ and $\{A,B \} = AB + BA$, respectively, and also defined 
\begin{align}
    \mathcal{D}_C \equiv 
    \begin{cases}
     k^2 - \frac{1}{4} \partial_x^2 \\
     \cancel{k} 
    \end{cases}, ~~~~\mathcal{D}_K \equiv
    \begin{cases}
     ik \cdot \partial_x ~~~~&(\mathrm{bosons}) \\
    \frac{i}{2} \cancel{\partial}_x ~~~~&(\mathrm{fermions})
    \end{cases}.
\end{align}
In Eqs.~\eqref{eq:KBeq_0} and \eqref{eq:KBeq_1}, we have used the shorthand notation
\begin{align}
  e^{-i\diamond}\{A,B\}
  &\equiv
  e^{-i\diamond}(A,B)+e^{-i\diamond}(B,A),
  \notag \\
  e^{-i\diamond}[A,B]
  &\equiv
  e^{-i\diamond}(A,B)-e^{-i\diamond}(B,A).
\end{align}
The Hermitian conjugate operation for the propagators is summarized in appendix~\ref{sec:hermiticity}.

For a (single-flavor) bosonic system, Eq.~\eqref{eq:KBeq_0} plays the role of the constraint equation.
It constrains the solution of the Wightman functions and determines the on-shell condition.
The Hermitian part, Eq.~\eqref{eq:KBeq_1}, plays the role of the kinetic equation and determines the evolution of the on-shell solution.
For a (single-flavor) fermionic system, the analogous constraint and kinetic equations are obtained after decomposing the spinor structure of Eq.~\eqref{eq:Wightman1} and then taking the Hermitian and anti-Hermitian parts.

In conventional derivations~\cite{Kainulainen:2001cn,Prokopec:2003pj,Kainulainen:2002th,Kainulainen:2021oqs}, one first solves the constraint equation, usually neglecting the self-energy corrections or including only part of their effects.
This admits a homogeneous solution proportional to a delta function that imposes the on-shell condition.
The Boltzmann equation is then obtained by substituting this on-shell solution into the kinetic equation.
In contrast, in section~\ref{sec:solution}, we solve Eqs.~\eqref{eq:propag} and \eqref{eq:Wightman1}, or equivalently Eqs.~\eqref{eq:propag} and \eqref{eq:Wightman2}, directly and order by order in the derivative expansion in the presence of the self-energy.
We then apply the approximations relevant for EWBG to the solutions and identify the on-shell part of the Wightman functions.
As discussed in section~\ref{sec:Boltzmann}, the Boltzmann equation is obtained by substituting this on-shell part into Eq.~\eqref{eq:KBeq_1}.

\subsection{Local basis transformation \label{sec:basis}}

In this subsection, we introduce a local basis in which the spacetime-dependent mass term is diagonalized at each spacetime point.
This local mass eigenstate basis, which we hereafter refer to as the local mass basis, is essentially the same as that commonly used in calculations of EWBG~\cite{Kainulainen:2001cn,Prokopec:2003pj}.
We use this basis to identify the on-shell part of the solutions and thereby derive the Boltzmann equation.

We start from a flavor basis in which the kinetic term is diagonalized in flavor space, while the mass matrix is generally non-diagonal and depends on the spacetime.
The local mass basis is defined by a spacetime-dependent unitary transformation that diagonalizes the mass matrix at each point.
Because the transformation to the local mass basis is spacetime dependent, derivatives acting on the rotation matrices generate additional terms in the Wigner representation.
These terms can be regarded as connection terms associated with the spacetime-dependent rotation in flavor space.

Bosons and fermions in the local mass basis are given by those in the flavor basis as 
\begin{align}
  &\hat{\phi}(u) = U_b(u) \phi(u), \notag \\
  &\hat{\psi}(u) = U_f(u) \psi(u),
\end{align} 
where $U_b$ is a unitary matrix in flavor space, while $U_f$ also acts on spinor space and is decomposed as
\begin{align}
  U_f(u) = P_L \otimes U_L(u) + P_R \otimes U_R(u),
\end{align}
where $P_{L,R} \equiv (1 \mp \gamma^5)/2$ and $U_{L,R}$ are unitary.

The mass term is diagonalized by these unitary matrices.
For bosons, we have
\begin{align}
  \hat{M}^2(u) = U_b(u) M^2(u) U_b^\dagger (u),
  \label{eq:massdiagonalize_boson}
\end{align}
where
\begin{align}
  \hat{M}^2 = \mathrm{diag}(m_1^2,\cdots,m_n^2),~~~~(m_i^2 \in \mathbb{R}).
\end{align}
For fermions, the mass term $-\overline{\psi_L} (M_0 + i M_5) \psi_R + \mathrm{h.c.}$ is diagonalized as
\begin{align}
  \hat{M}_0 + i \hat{M}_5 = U_L (M_0 + i M_5) U_R^\dagger = \mathrm{diag}(m_1,\cdots,m_n),~~~~(m_i \in \mathbb{C}),
  \label{eq:massdiagonalize_fermion}
\end{align}
or equivalently we can write
\begin{align}
  &\hat{M}_0 = \mathrm{diag}(m_{R1},\cdots,m_{Rn})~~~~(m_{Ri} \in \mathbb{R}), \notag \\
  &\hat{M}_5 = \mathrm{diag}(m_{I1},\cdots,m_{In})~~~~~(m_{Ii} \in \mathbb{R}).
\end{align}

In this basis, not only the mass matrix but also all propagators and self-energy functions are transformed.
Quantities in this basis are denoted by hats.
To treat the bosonic and fermionic cases in a unified manner, we let $X$ denote either $\Delta$ or $\Pi$ and set $U=U_b$ for bosons, while $X$ denotes either $\gamma^0 S$ or $\Sigma\gamma^0$ and $U=\overline{U_f}^\dagger$ for fermions, where $\overline{U_f} \equiv \gamma^0 U_f^\dagger \gamma^0$.
We then have 
\begin{align}
  \hat{X}(k,x) &=e^{-i \diamond_3} \Big( U (x) X (k,x) U^\dagger (x) \Big),
\end{align}
where the operator $\diamond_3$ acts on the three factors in the product, and its $n$-th power is defined by
\begin{align}
  \diamond_3^n \Big( A(x) B(k,x) C(x) \Big) \equiv \frac{1}{2^n}\big( \partial^A_x \cdot \partial^B_k - \partial^B_k \cdot \partial^C_x \big)^n  A(x) B(k,x) C(x) .
\end{align}
Expanding the exponential to first order in derivatives, we obtain
\begin{align}
  \hat{X}(k,x) &= U(x) X(k,x) U^\dagger(x) - \frac{i}{2} \Big\{ \big(\partial_x U(x) \big)U^\dagger(x) , \, \partial_{k} \big( U(x) X(k,x) U^\dagger (x) \big) \Big\} + O(\partial^2_x).
  \label{eq:masstoflavor}
\end{align}

Even after diagonalizing the mass matrix, the transformation to the mass basis remains undetermined up to independent phase rotations of the mass eigenstates, which leave the diagonal mass matrix unchanged.
It is convenient to choose the basis where the diagonal part of $(\partial_x U) U^\dagger$ is removed~\cite{Kainulainen:2001cn,Prokopec:2003pj}.
For the bosonic case, under the diagonal rotation $U_b \to \mathrm{diag}(e^{i\varphi_1}, \cdots, e^{i\varphi_n}) U_b$, which does not change $\hat{M}^2$, we have 
\begin{align}
  (\partial_x U_b) U^\dagger_b \to 
  \begin{cases}
    [(\partial_x U_b) U^\dagger_b]_{ii} + i \partial_x \varphi_i, ~~(i = j) \\
    [(\partial_x U_b) U^\dagger_b]_{ij} e^{i (\varphi_i - \varphi_j)},~~(i \neq j)
  \end{cases}.
\end{align}
Therefore, the diagonal part of $(\partial_x U_b) U^\dagger_b$ is rephasing dependent and can be removed by choosing $i \partial_x \varphi_i = - [(\partial_x U_b) U^\dagger_b]_{ii}$.
Similarly, for fermions, under $U_L \to \mathrm{diag}(e^{i\varphi_{1}^L}, \cdots, e^{i\varphi_n^L}) U_L$ and $U_R \to \mathrm{diag}(e^{i\varphi_{1}^R}, \cdots, e^{i\varphi_n^R}) U_R$, we obtain
\begin{align}
  (\partial_x U_f) U^\dagger_f &= (\partial_x U_L) U^\dagger_L P_L + (\partial_x U_R) U^\dagger_R P_R \notag \\
  &\to 
  \begin{cases}
    \Big( [(\partial_x U_L) U^\dagger_L]_{ii} + i \partial_x \varphi_i^L \Big) P_L + \Big( [(\partial_x U_R) U^\dagger_R]_{ii} + i \partial_x \varphi_i^R \Big) P_R, ~~(i = j) \\
    [(\partial_x U_L) U^\dagger_L]_{ij} e^{i (\varphi_i^L - \varphi_j^L)}P_L + [(\partial_x U_R) U^\dagger_R]_{ij} e^{i (\varphi_i^R - \varphi_j^R)}P_R ,~~~~~~~(i \neq j)
  \end{cases}.
  \label{eq:fermion_connection_trans}
\end{align}
Therefore, by taking $i \partial_x \varphi_i^L = - [(\partial_x U_L) U^\dagger_L]_{ii}$ and $i \partial_x \varphi_i^R = - [(\partial_x U_R) U^\dagger_R]_{ii}$, the diagonal part of $(\partial_x U_f) U^\dagger_f$ can be removed.
We note that, in contrast to the bosonic case, the phase convention of the fermion mass eigenvalue changes as
\begin{align}
  |m_i| e^{i \theta_i} \to |m_i| e^{i (\theta_i + \varphi_i^L - \varphi_i^R)} ,
  \label{eq:mass_replace}
\end{align}
where $|m_i| e^{i \theta_i} \equiv m_{Ri} + i m_{Ii}$.

\section{Conditions to obtain the Boltzmann equation for electroweak baryogenesis \label{sec:approx}}

In this section, we clarify the approximations and assumptions to obtain the Boltzmann equation from the Kadanoff--Baym equations for EWBG.
We consider the standard picture~\cite{Cline:1997vk,Cline:2000nw,Kainulainen:2001cn,Kainulainen:2002th,Prokopec:2003pj,Prokopec:2004ic,Fromme:2006wx,Cline:2020jre,Kainulainen:2021oqs,Kainulainen:2024qpm}, in which bubbles of the broken phase expand during a first-order electroweak phase transition.
Around the bubble wall, the vacuum expectation values of scalar fields vary in spacetime, and the mass matrices of particles become spacetime dependent.
Particles propagating through this slowly varying background are then described, after suitable approximations, as quasiparticles whose distribution functions obey Boltzmann equations.
The purpose of this section is to state the assumptions under which this quasiparticle Boltzmann description is extracted from the Kadanoff--Baym equations.

\subsection{Spacetime-dependent background and the derivative expansion}

In the Wigner representation, the mass term $\mathcal{M}(x)$ is a function of the average coordinate $x$, while the self-energy functions generally also depend on the momentum $k$ and the plasma four-velocity $u$:
\begin{align}
  g^{r,a,\lambda} = g^{r,a,\lambda}(k,u,x).
\end{align}
In the EWBG setup, the $x$ dependence of the system is mainly induced by the spacetime-dependent vacuum expectation value as a background around the bubble wall.
In principle, this background can also induce an $x$ dependence in the self-energy functions through field-dependent masses and local plasma properties.
In solving the Kadanoff--Baym equations and deriving the solutions, however, we neglect gradient terms in which $x$-derivatives act on the self-energy functions.
With this approximation, the spacetime dependence relevant for the first-order solutions enters through the mass term induced by the vacuum expectation value.

In the presence of the spacetime-dependent mass term, we cannot fully transform the system to momentum space in the usual way.
To solve the equations, we assume that the gradients with respect to $x$ are much smaller than the microscopic scale characterized by the momentum $k$.
This approximation is schematically denoted by $k \gg \partial_x$.
Due to this hierarchy, higher-order terms in the expansion of $e^{-i\diamond}$ (see Eq.~\eqref{eq:exp_expand}) are suppressed by additional powers of $\partial_x/k$, and the expansion can be truncated at the required order.
Similarly, we expand $\mathcal{D}$ and the propagators order by order as
\begin{align}
  \mathcal{D} = \sum_{n=0}^{(3+\eta)/2} \mathcal{D}^{[n]}, ~~~~
  G = \sum_{n=0}^\infty G^{[n]},
\end{align}
where $\eta = +1$ $(-1)$ for bosons (fermions), and the $n$-th order in the derivative expansion is denoted by $[n]$.
Since the combination $\hbar \, \partial_x \cdot \partial_k$ is dimensionless, the derivative expansion corresponds to an $\hbar$ expansion.

\subsection{Planar-wall approximation, steady-wall assumption, and spinor decomposition \label{sec:spindecompose}}

We introduce the planar-wall approximation and the steady-wall assumption, and the spinor decomposition for fermions.

The planar-wall approximation is justified when the bubble radius is much larger than the wall width and the microscopic length scales relevant for particle propagation.
We choose the $z$ axis to be normal to the wall and denote the coordinates parallel to the wall by $\bm{x}_\parallel$.
In the planar-wall approximation, the wall background is independent of $\bm{x}_\parallel$.

We also assume that, after nucleation and the initial stage of acceleration, the bubble wall reaches a steady propagation with a constant velocity $v_w$ in a frame in which the plasma far in front of the wall is at rest.
In this regime, the wall profile does not change its shape during the propagation.
We work in the wall frame, namely the frame comoving with the bubble wall.
In the wall frame, the mass term is static and depends only on the coordinate $z$, $\mathcal{M}(x) =\mathcal{M}(z)$.
The wall background remains translationally invariant along the wall plane.
As a result, in the propagation through the wall background, $\bm{k}_\parallel$ is conserved and can be used to label each mode.
If we neglect the wall effects on the plasma flow, the plasma four-velocity in the wall frame is
$u^\mu=u_w^\mu=\gamma_w(1,0,0,-v_w)$, where $\gamma_w=1/\sqrt{1-v_w^2}$.

For fermions, it is useful to analyze each mode in the boosted frame where the parallel momentum $\bm{k}_\parallel$ vanishes.
We denote quantities in this frame by a tilde.
The corresponding boost acts only in the directions parallel to the wall and gives
\begin{align}
  \tilde{\bm{k}}_\parallel = 0,
  \qquad
  \tilde{k}_z = k_z,
  \qquad
  \tilde{k}^0 =
  \mathrm{sign}(k^0)\sqrt{(k^0)^2-|\bm{k}_\parallel|^2} \equiv
  \gamma_\parallel^{-1}k^0.
\end{align}
We write the plasma four-velocity in the boosted frame as $\tilde{u}^\mu$.
The explicit form of the Lorentz boost is given in appendix~\ref{sec:boost}.

In the boosted frame, the fermionic kinetic operator and the mass term depend only on $\tilde{k}^0$, $k_z$, and $z$.
We further assume that the plasma effects included in the self-energy preserve the spin projection along the $z$ direction in the boosted frame.
The corresponding spin operator is
\begin{align}
  S_z \equiv \gamma^0 \gamma^3 \gamma^5 .
\end{align}
The operator $S_z$ commutes with the kinetic and mass operators and the self-energy in the boosted frame.
Therefore, the spinor structure of the propagators can be decomposed into two independent sectors corresponding to the spin projections $s=\pm$ along the $z$ direction~\cite{Kainulainen:2001cn,Kainulainen:2002th}.
The two spin projections $s=\pm$ span a two-dimensional spin space.
Within each spin sector, the remaining two-dimensional space is spanned by the left- and right-handed chiral components, which we refer to as the chiral space.
The four-component spinor space is thus written as the product of the spin and chiral spaces.

In the Weyl representation for the Dirac matrices, the spinor decomposition is explicitly given as follows.
The spin operator is written as
\begin{align}
  S_z = \sigma^3 \otimes I,
\end{align}
where the third Pauli matrix $\sigma^3$ acts in the spin space and $I$ is the identity matrix in the chiral space.
The projection operator onto the spin $s$ in the spin space is then defined by
\begin{align}
  P_s \equiv \frac{1}{2}(1+s\sigma^3).
\end{align}
Accordingly, the fermionic propagators and the self-energy in the boosted frame are decomposed as
\begin{align}
  - i \gamma^0 S^{r,a,\lambda}
  &=
  \sum_{s=\pm} P_s \otimes \frac{1}{2}
  \left( g_{0s}^{r,a,\lambda} + \sum_{m=1}^3 \sigma^m g_{ms}^{r,a,\lambda} \right), \notag \\
  \Sigma^{r,a,\lambda}\gamma^0 &= \sum_{s=\pm} P_s \otimes
  \left( a_{0s}^{r,a,\lambda} + \sum_{m=1}^3 \sigma^m a_{ms}^{r,a,\lambda} \right).
  \label{eq:spinordecompose}
\end{align}
The coefficients $g_{0s}^{r,a,\lambda}$ and $a_{0s}^{r,a,\lambda}$ multiply the identity matrix in the chiral space, while $g_{ms}^{r,a,\lambda}$ and $a_{ms}^{r,a,\lambda}$ multiply the Pauli matrices $\sigma^m$.
All coefficients $g_{0s}^{r,a,\lambda}$, $g_{ms}^{r,a,\lambda}$, $a_{0s}^{r,a,\lambda}$, and $a_{ms}^{r,a,\lambda}$ are matrices in flavor space.

We emphasize that this spinor decomposition is defined in the boosted frame.
The self-energy coefficients are functions of the boosted momentum, the boosted plasma velocity, and the coordinate $z$, e.g. $a_{0s}^{r,a,\lambda}=a_{0s}^{r,a,\lambda}(\tilde{k},\tilde{u},z)$.
Their explicit forms are not needed in the following derivation; explicit expressions can be found, for example, in ref.~\cite{Kainulainen:2021oqs}.
After obtaining the solution in the boosted frame, we transform the final fermionic results back to the original wall frame.

\subsection{Quasiparticle approximation}

We explain the quasiparticle approximation used in the following sections.
In the presence of interactions, the spectral function generally has a peak with a finite width rather than a delta-function peak.
Nevertheless, if the width is sufficiently small, the spectral function is sharply peaked near the energies determined by the dispersion relation.
We denote this condition generically by
\begin{align}
  \Omega^2(k,x)=0 .
\end{align}
We call $\Omega^2(k,x)$ the shell function and refer to the equation $\Omega^2(k,x)=0$ as the on-shell condition for a quasiparticle.
We also refer to the set of real variables $(k,x)$ satisfying this condition as the quasiparticle shell.
For fixed $(\bm{k},x)$, a solution of the on-shell condition belongs to a positive-energy branch if $k^0>0$ and to a negative-energy branch if $k^0<0$.
The part of the solution proportional to $\delta(\Omega^2)$ is referred to as the on-shell part.

In this work, the quasiparticle approximation means the following procedure.
When the spectral function contains a Breit--Wigner-type structure around the on-shell condition $\Omega^2(k,x)=0$, we take the narrow-width limit (NWL) of this structure and replace it by a delta function $\delta(\Omega^2)$.
The resulting on-shell part is then used as the quasiparticle contribution to the spectral and Wightman functions.

For illustration, let us first consider a single bosonic degree of freedom.
The same argument applies to the multi-flavor case, where flavor mixing effects discussed in section~\ref{sec:mixing} are neglected.
At zeroth order in the derivative expansion, we will encounter retarded and advanced propagators of the form
\begin{align}
  \Delta^{r,a} (k,x) \simeq \frac{1}{\Omega^2(k,x) \pm i \Pi_{{\mathcal A}}(k)} ,
\end{align}
where the equation $\Omega^2(k,x)=0$ gives the on-shell condition when the absorptive part is neglected.
The absorptive part $\Pi_{\mathcal A}$ shifts the pole away from the real axis and controls the width of the Breit--Wigner peak around $\Omega^2(k,x)=0$.
The corresponding spectral function is
\begin{align}
  \mathscr{A} (k,x) \simeq \frac{\Pi_{{\mathcal A}}(k)}{\Omega^4(k,x) + \Pi_{{\mathcal A}}^2(k)} .
\end{align}
When the width is sufficiently small, the spectral function is sharply peaked near the on-shell condition $\Omega^2(k,x)=0$.
The NWL then gives
\begin{align}
  \mathscr{A} (k,x)
  \xrightarrow{\rm NWL}
  \pi \, \mathrm{sign}(k^0) \, \delta\big( \Omega^2 (k,x) \big).
\end{align}
Equivalently, by rewriting $\delta(\Omega^2)$ in terms of the roots of the on-shell condition, the spectral function is expressed as a sum over positive- and negative-energy branches.
We note that the Hermitian part of the self-energy can modify the shell function and hence the dispersion relation.

For fermions, the above procedure can be applied after the spinor decomposition.
In the boosted frame introduced in section~\ref{sec:spindecompose}, the two spin sectors can be treated independently under the assumption that the self-energy preserves the spin projection along the $z$ direction.
At zeroth order in the derivative expansion, the spectral function for a spin-$s$ mode takes the schematic Breit--Wigner-type structure
\begin{align}
  P_s \big(\gamma^0 \mathcal{A}(k,x) \big) \simeq
  P_s \otimes
  \frac{\Gamma_s(k,x)}{\Omega_s^4(k,x)+\Gamma_s^2(k,x)}
  \mathcal{N}_s(k,x),
\end{align}
where $\mathcal{N}_s$ is a $2\times 2$ matrix in the chiral space.
The function $\Omega_s^2(k,x)$ is the spin-dependent shell function, and the equation $\Omega_s^2(k,x)=0$ gives the corresponding on-shell condition when the absorptive part is neglected.
The quantity $\Gamma_s$ is determined by the anti-Hermitian part of the self-energy and controls the width of the Breit--Wigner peak around $\Omega_s^2(k,x)=0$.
When the width is sufficiently small, the NWL gives
\begin{align}
  P_s \big(\gamma^0 \mathcal{A}(k,x) \big)
  \xrightarrow{\rm NWL}
  P_s \otimes
  \pi \, \mathrm{sign}(\tilde{k}^0) \,
  \delta\big( \Omega_s^2(k,x) \big)
  \mathcal{N}_s(k,x).
\end{align}
The roots of the spin-dependent on-shell condition define the positive- and negative-energy branches in each spin sector\footnote{The Hermitian part of the fermionic self-energy modifies the on-shell condition and can lead to additional real roots. 
For an explicit analysis of the resulting dispersion relations, see ref.~\cite{Kainulainen:2021oqs}.}.
A more explicit realization of this structure will be given in section~\ref{sec:spectral}.

The sign factors appearing in the NWL deserve a comment.
They arise when $\Pi_{\mathcal{A}}$ for bosons and $\Gamma_s$ for fermions are positive on the respective positive-energy branches and have the appropriate odd behavior under energy reversal.
For bosons, the latter condition corresponds to
\begin{align}
  \Pi_{\mathcal A}(k^0) = - \Pi_{\mathcal A}(-k^0),
\end{align}
which follows from the CP invariance of the self-energy (see appendix~\ref{sec:hermiticity}).
For fermions, the CP transformation of the self-energy maps the spin label $s$ to $-s$.
Therefore, the simple sign factor in each spin sector is obtained under additional assumptions on the spin and chiral structure of the absorptive part.
In particular, we assume that the absorptive part relevant for $\Gamma_s$ is independent of the spin label $s$ and dominated by the scalar component in the chiral space, $\overline{a}_{0,\mathcal A} \gg \overline{a}_{m,\mathcal A}$, so that $\Gamma_s= \mathcal{O}(\overline{a}_{0,\mathcal A})$ (cf. Eq.~\eqref{eq:fermion_omega_gamma}).
The NWL is taken by making $\overline{a}_{0,\mathcal A}$ small while maintaining this hierarchy.
Furthermore, for later use, we assume that momentum differentiation does not alter this hierarchy, so that $\partial_{k_z}\overline{a}_{0,\mathcal A}=\mathcal{O}(\overline{a}_{0,\mathcal A})$ and $\partial_{k_z}\overline{a}_{m,\mathcal A}=\mathcal{O}(\overline{a}_{m,\mathcal A})$ in this limit.
Under the assumptions on the spin and chiral structure stated above, CP invariance implies the appropriate odd behavior of $\Gamma_s$.
Together with its positivity on the positive-energy branch, this gives the factor $\mathrm{sign}(\tilde{k}^0)$ in each spin sector in the NWL.
If the absorptive part has sizable spin dependence or a nontrivial chiral structure, the spectral function does not necessarily take the simple form $\mathrm{sign}(\tilde{k}^0)\delta(\Omega_s^2)$ in each spin sector.
In such cases, a more general quasiparticle approximation would be required.
We do not discuss this possibility further and use the quasiparticle approximation in the form written above.

The NWL should not be confused with setting the self-energy strictly to zero everywhere.
The latter would remove the interaction effects responsible for the collision term.
Instead, in the quasiparticle approximation, the Breit--Wigner-type structure is replaced by a delta function near $\Omega^2 (k,x) =0$, while the collision term generated by the self-energy is kept.

As we will see in section~\ref{sec:spectral}, beyond zeroth order in the derivative expansion, the shell function can receive corrections from the gradient of the wall.
In particular, for fermions, corrections at first order in the derivative expansion to the spectral function can generate terms proportional to $\delta'(\Omega^2)$ in the NWL.
These terms can be interpreted as shifts of the quasiparticle pole position relative to its zeroth-order value.

\subsection{Perturbation for flavor mixing \label{sec:mixing}}

We explain how flavor-dependent interactions are treated in the following analysis.
In general, the self-energy can contain both flavor-universal and flavor-dependent parts.
We decompose it as
\begin{align}
  (g^{r,a,\lambda})_{ij}
  =
  \overline{g}^{r,a,\lambda} \, \delta_{ij}
  +
  (\delta g^{r,a,\lambda})_{ij},
  \label{eq:flavorperturbation}
\end{align}
where the flavor indices $i,j$ have been explicitly shown.
The first term on the right-hand side is the flavor-universal part, while the second term is the flavor-dependent part.
Since the flavor-universal part is proportional to the identity matrix in flavor space, it remains diagonal in the local mass basis.

In section~\ref{sec:systematic}, we assume that the flavor-dependent part $\delta g^{r,a,\lambda}$ is small compared with the flavor-universal part $\overline{g}^{r,a,\lambda}$.
Under this assumption, the quasiparticle shell is determined by the diagonalized mass matrix and the flavor-universal self-energy.
The flavor-dependent part inducing flavor mixing and flavor-oscillation effects could be included perturbatively.
In deriving the Boltzmann equation in this work, for simplicity, we neglect the flavor-dependent part and retain only the flavor-universal part of the self-energy.
We refer to this treatment as the flavor-universal approximation.
Under this approximation, the solutions at zeroth order in the derivative expansion are diagonal in flavor space.

Even after this simplification, at first order in the derivative expansion, the solutions in the local mass basis are, in general, not diagonal in flavor space.
This is because the spacetime dependence of the local basis generates connection terms, which can induce off-diagonal components in the spectral and Wightman functions.
In this work, however, we aim to derive the ordinary semiclassical Boltzmann equation for the flavor-diagonal quasiparticles.
We therefore restrict ourselves to the diagonal components in the local mass basis and do not keep the off-diagonal components.
This point is discussed further in section~\ref{sec:spectral}.

\subsection{Self-energy decomposition and fluid ansatz \label{sec:selfenergydecompose}}

We adopt the following prescription: the Wightman self-energy $g^\lambda$ is separated into a contribution that specifies the on-shell part of the Wightman functions $G^\lambda$ and a remaining contribution to the collision term.
This decomposition also provides a natural way to introduce the statistical function and connect it to the fluid ansatz used in EWBG calculations.

In our setup, as we will see in section~\ref{sec:solution}, the Wightman functions $G^\lambda$ are obtained algebraically in terms of the spacetime-dependent mass term and the Wightman self-energy $g^\lambda$.
Here $g^\lambda$ is the two-point function of the composite operators coupled to the particle of interest.
The functions $G^\lambda$ should therefore be interpreted as dressed Wightman functions, in which the effects of the mass term and the self-energy corrections are incorporated order by order in the derivative expansion.
In this formulation, the statistical information of the system is introduced through the Wightman self-energy $g^\lambda$.
Based on this observation, we decompose the Wightman self-energy as\footnote{
    This decomposition should be understood as a prescription for identifying the on-shell Wightman functions.
    As explained below, at zeroth order in the derivative expansion, the local Kubo--Martin--Schwinger relation fixes $n^\lambda=n^\lambda_{\mathrm{leq}}$, while the remaining contribution vanishes, $\delta g_{\mathrm{coll}}^{[0]}=0$.
}
\begin{align}
  g^\lambda
  \equiv
  g^\lambda_{\mathrm{stat}}
  +
  \delta g_{\mathrm{coll}},
  \label{eq:selfenergy_decomposition}
\end{align}
where
\begin{align}
  g^\lambda_{\mathrm{stat}}
  =
  \frac{2\eta}{i} n^\lambda(k,x) g_{\mathcal A}.
  \label{eq:g_stat_def_working}
\end{align}
Here $n^\lambda(k,x)$ is a non-singular statistical function and satisfies 
\begin{align}
    n^{>} = n^{<} +\eta .
\end{align}
We assume that $n^\lambda$ is a c-number.

The role of $g^\lambda_{\mathrm{stat}}$ is to generate the on-shell part of the dressed Wightman functions $G^\lambda$.
As we will derive in section~\ref{sec:Wightman}, by substituting $g^\lambda = g^\lambda_{\mathrm{stat}}$ into the solutions for $G^\lambda$ at zeroth order in the derivative expansion, we obtain the Wightman functions as $G^{\lambda,[0]} \propto n^\lambda \mathscr{A}^{[0]}$.
After taking the quasiparticle approximation and projecting the on-shell part onto a particular branch, $n^\lambda$ is related to $f_i$, the distribution function for quasiparticle species $i$.
Therefore, the non-singular function $n^\lambda$ carries the statistical information of the on-shell excitation.
At first order in the derivative expansion, the same prescription generates the corresponding on-shell terms involving gradients of $n^\lambda$ and of the spacetime-dependent mass term.

In contrast to $g^\lambda_{\mathrm{stat}}$, $\delta g_{\mathrm{coll}}$ denotes the remainder of $g^\lambda$ after subtracting $g^\lambda_{\mathrm{stat}}$.
It depends on the distributions of the bath or internal particles involved in scattering and decay processes.
Therefore, it should not be included in the on-shell part of the Wightman functions.
Instead, it is kept in the self-energy entering the collision term.
In section~\ref{sec:Boltzmann}, the collision term will be obtained by inserting the Wightman self-energy $g^\lambda$ into the right-hand side of Eq.~\eqref{eq:KBeq_1}.
The above decomposition is mainly used to identify the on-shell part of the Wightman functions and to specify which contributions to the collision term are retained at the order considered in the derivative expansion.

We also discuss the relation between the decomposition in Eq.~\eqref{eq:selfenergy_decomposition} and the fluid ansatz for the distribution function.
At zeroth order in the derivative expansion, we assume local thermal equilibrium on the wall background, with the temperature and plasma four-velocity taken to be independent of $z$.
In the wall frame, $u^\mu=u_w^\mu$, and the distribution function takes the form
\begin{align}
  f_i(k,x) \Big|_{n=0} = f_{i,0w} (k,x) = \frac{1}{e^{\beta[\gamma_w (\omega_i + v_w k_z) ]} - \eta},
\end{align}
where $\beta$ is the inverse temperature and $\omega_i$ is the $z$-dependent energy.
Consistently with this local-thermal-equilibrium ansatz, at zeroth order in the derivative expansion, the Wightman self-energy $g^\lambda$ satisfies the local Kubo--Martin--Schwinger (KMS) relation
\begin{align}
  g^\lambda = g_{\rm stat}^{\lambda} = \frac{2\eta}{i}
  n^{\lambda}_{\rm leq}
  g_{\mathcal A},
  \label{eq:gcut_gstat_leq}
\end{align}
where
\begin{align}
  n^{<}_{\rm leq}
  =
  \frac{1}{e^{\beta k \cdot u_w } - \eta},
  \qquad
  n^{>}_{\rm leq}
  =
  n^{<}_{\rm leq}+\eta .
  \label{eq:n_leq}
\end{align}
For fermions, the calculation is performed in the boosted frame, where the same ansatz is expressed in terms of $\tilde{k} \cdot \tilde{u}_w = k \cdot u_w$.

Beyond zeroth order, terms at first order in the derivative expansion drive the distribution $f_i$ away from local thermal equilibrium. 
We can parametrize this departure by the fluid ansatz commonly used in EWBG calculations~\cite{Huet:1995sh,Cline:2000nw,Kainulainen:2002th,Prokopec:2003pj,Prokopec:2004ic,Fromme:2006wx,Cline:2020jre,Kainulainen:2021oqs,Kainulainen:2024qpm},
\begin{align}
  f_i(k,x) \simeq f_{i,0w}(k,x) - \beta \mu_i f_{i,0w}^\prime + \delta f_i(k,x),
  \label{eq:distribution_wall_perturbation}
\end{align}
where $\mu_i$ is the chemical potential and $f_{i,0w}^\prime$ denotes the derivative of $f_{i,0w}$ with respect to its argument $\beta\gamma_w(\omega_i+v_wk_z)$.
The second term on the right-hand side describes the local kinetic-equilibrium perturbation, while the third describes the deviation from local kinetic equilibrium\footnote{
    If the wall velocity is very close to the speed of light, this assumption may break down.
    For a more general approach, see refs.~\cite{Laurent:2020gpg,Dorsch:2021ubz,Laurent:2022jrs}.
}.
Since these departures from local thermal equilibrium are induced by the gradient of the wall, $\beta\mu_i$ and $\delta f_i$ start at first order in the derivative expansion.
Correspondingly, $n^\lambda(k,x)$ is promoted from the local-thermal-equilibrium function $n^\lambda_{\rm leq}$ to a general statistical function.
Away from local thermal equilibrium, the local KMS relation in Eq.~\eqref{eq:gcut_gstat_leq} no longer holds.
We therefore use the more general decomposition in Eq.~\eqref{eq:selfenergy_decomposition}: the part proportional to $n^\lambda g_{\mathcal A}$ specifies the on-shell part, while the remaining part contributes to the collision term.
Since the latter part vanishes in local thermal equilibrium at zeroth order, its leading contribution starts at first order in the derivative expansion:
\begin{align}
  \delta g_{\mathrm{coll}}
  =
  \delta g_{\mathrm{coll}}^{[1]}+\cdots .
  \label{eq:delta_g_order_counting}
\end{align}

\section{Analytic solution in the derivative expansion \label{sec:solution}}

In this section, using the derivative expansion, we derive the solutions of the Kadanoff--Baym equations in Eqs.~\eqref{eq:propag} and \eqref{eq:Wightman2} up to first order in the derivative expansion.
We first show the solutions without any additional assumption other than the derivative expansion.
At this stage, we work in the flavor basis.
The solutions in the local mass basis are introduced by applying the local basis transformation in the Wigner space.

By solving Eqs.~\eqref{eq:propag} and \eqref{eq:Wightman2} at zeroth order, we obtain
\begin{align}
  &G^{r,a,[0]} = \alpha_\mp^{-1}, \notag \\
  &G^{\lambda,[0]} = \alpha_-^{-1} g^\lambda \alpha_+^{-1},
  \label{eq:sol_n0}
\end{align}
where 
\begin{align}
  \alpha_{\pm} \equiv \mathcal{D}^{[0]} - (\mathcal{M} + g_{\mathcal{H}} \pm i g_{\mathcal{A}}).
  \label{eq:def_alpha}
\end{align}
By substituting $G^{r,a,[0]}$ into the definition of the spectral function given in Eq.~\eqref{eq:propag3}, at zeroth order we obtain 
\begin{align}
  \mathscr{A}^{[0]} = \alpha_-^{-1} g_{\mathcal{A}} \alpha_+^{-1} = \alpha_+^{-1} g_{\mathcal{A}} \alpha_-^{-1} = \frac{i}{2}(\alpha_-^{-1} - \alpha_+^{-1}).
  \label{eq:n0_spectral}
\end{align}
At finite width, the absorptive part of the self-energy prevents $\alpha_\pm(k,x)$ from having zero eigenvalues for real $k^0$.
Thus, for a set of real variables $(k,x)$, including those on the quasiparticle shell, the matrices $\alpha_\pm(k,x)$ are invertible,
\begin{align}
  \det\alpha_\pm(k^0,\bm{k},x)\neq 0
  \qquad
  (k^0\in\mathbb{R}),
\end{align}
as long as the modes under consideration have a finite absorptive part.
Therefore, at finite width, the solutions in Eq.~\eqref{eq:sol_n0} are regular particular solutions sourced by the self-energy.
The singular contribution supported on the quasiparticle shell appears only after taking the quasiparticle approximation.
In contrast, if one takes the collisionless limit from the beginning, where all self-energy functions vanish, the equations admit a homogeneous solution~\cite{Kainulainen:2001cn,Prokopec:2003pj,Kainulainen:2002th}.
\footnote{
    The effective finite-width behavior obtained from a finite-width ansatz can be reproduced by summing to all orders the gradient terms acting on the on-shell delta function in the collision term~\cite{Garbrecht:2011xw}.
}

By expanding Eqs.~\eqref{eq:propag} and \eqref{eq:Wightman2} at first order in the derivative expansion, we obtain 
\begin{align}
  &G^{r,a,[1]} = \alpha_\mp^{-1}  i \diamond (\alpha_\mp, \alpha_\mp^{-1}) , \notag \\
  &G^{\lambda,[1]} = G^{\lambda,[0]} \big( i \diamond(\alpha_+,\alpha_+^{-1}) \big) + \alpha_-^{-1} i \diamond (\alpha_-, G^{\lambda,[0]}) - \alpha_-^{-1} \big(i \diamond (g^\lambda, \alpha_+^{-1}) \big).
  \label{eq:sol_n1}
\end{align}
By substituting $G^{r,a,[1]}$ into the definition of the spectral function in Eq.~\eqref{eq:propag3}, we obtain the first-order contribution,
\begin{align}
  \mathscr{A}^{[1]} = \frac{i}{2} \big( \alpha_-^{-1} i \diamond (\alpha_-, \alpha_-^{-1}) - \alpha_+^{-1} i \diamond (\alpha_+, \alpha_+^{-1}) \big).
  \label{eq:n1_spectral}
\end{align}

For later calculations, it is convenient to rewrite the solutions in the local mass basis introduced in section~\ref{sec:basis}.
Since the local basis transformation is spacetime dependent, the Wigner-space transformation generates connection terms.
We define the following local-basis components:
\begin{align}
  \alpha_{\phi, \pm} \equiv U_b \alpha_\pm U_b^\dagger = k^2 - \big( \hat{M}^2 + U_b \Pi_{\mathcal{H}} U_b^\dagger \pm i U_b \Pi_{\mathcal{A}} U_b^\dagger \big),
  \label{eq:def_alpha_bos}
\end{align} 
for bosons, and 
\begin{align}
  \alpha_{\psi, \pm} \equiv \overline{U_f}^\dagger \alpha_\pm \gamma^0  \overline{U_f} = \cancel{k} \gamma^0 - \big( \hat{M}_0 \gamma^0 + i \gamma^5 \gamma^0 \hat{M}_5 + \overline{U_f}^\dagger \Sigma_{\mathcal{H}} \gamma^0  \overline{U_f} \pm i \overline{U_f}^\dagger \Sigma_{\mathcal{A}} \gamma^0  \overline{U_f} \big),
  \label{eq:def_alpha_fer}
\end{align} 
for fermions.
Applying the basis transformation given in Eq.~\eqref{eq:masstoflavor} to Eqs.~\eqref{eq:sol_n0}, \eqref{eq:n0_spectral}, \eqref{eq:sol_n1}, and \eqref{eq:n1_spectral}, we obtain the spectral and Wightman functions at each order in the local mass basis:
\begin{align}
    \text{Bosons}:~~~&\hat{\mathscr{A}}_\phi^{[0]} = \frac{i}{2} (\alpha_{\phi, -}^{-1} - \alpha_{\phi, +}^{-1}), \notag \\
  &\hat{\mathscr{A}}_\phi^{[1]} = \frac{i}{2} \Big( \alpha_{\phi, -}^{-1} i \diamond_\Xi (\alpha_{\phi, -}, \alpha_{\phi, -}^{-1}) - \alpha_{\phi, +}^{-1} i \diamond_\Xi (\alpha_{\phi, +}, \alpha_{\phi, +}^{-1}) \Big) -\frac{i}{2} \big\{ \Xi, \partial_k \hat{\mathscr{A}}_\phi^{[0]} \big\}, \notag \\
  &\hat{\Delta}^{\lambda,[0]} = \alpha_{\phi, -}^{-1} U_b \Pi^\lambda U_b^\dagger \alpha_{\phi, +}^{-1}, \notag \\
  &\hat{\Delta}^{\lambda,[1]} = \hat{\Delta}^{\lambda,[0]}  i \diamond_\Xi (\alpha_{\phi, +}, \alpha_{\phi, +}^{-1}) + \alpha_{\phi, -}^{-1} i \diamond_\Xi (\alpha_{\phi, -}, \hat{\Delta}^{\lambda,[0]})  \notag \\
  &\qquad \quad - \alpha_{\phi, -}^{-1} i \diamond_\Xi (U_b \Pi^\lambda U_b^\dagger, \alpha_{\phi, +}^{-1}) -\frac{i}{2} \big\{ \Xi, \partial_k \hat{\Delta}^{\lambda, [0]} \big\}. \label{eq:sol_masseigenstate_bos} \\
  \text{Fermions}:~~~&\gamma^0 \hat{\mathscr{A}}_\psi^{[0]} = \frac{i}{2} (\alpha_{\psi, -}^{-1} - \alpha_{\psi, +}^{-1}), \notag \\
  &\gamma^0 \hat{\mathscr{A}}_\psi^{[1]} = \frac{i}{2} \Big( \alpha_{\psi, -}^{-1} i \diamond_\Xi (\alpha_{\psi, -}, \alpha_{\psi, -}^{-1}) - \alpha_{\psi, +}^{-1} i \diamond_\Xi (\alpha_{\psi, +}, \alpha_{\psi, +}^{-1}) \Big) -\frac{i}{2} \big\{ \Xi, \partial_k \gamma^0 \hat{\mathscr{A}}_\psi^{[0]} \big\}, \notag \\
  &\gamma^0 \hat{S}^{\lambda,[0]} = \alpha_{\psi, -}^{-1} \overline{U_f}^\dagger \Sigma^\lambda \gamma^0 \overline{U_f} \alpha_{\psi, +}^{-1}, \notag \\
  &\gamma^0 \hat{S}^{\lambda,[1]} = \gamma^0 \hat{S}^{\lambda,[0]}  i \diamond_\Xi (\alpha_{\psi, +}, \alpha_{\psi, +}^{-1})  + \alpha_{\psi, -}^{-1} i \diamond_\Xi (\alpha_{\psi, -}, \gamma^0 \hat{S}^{\lambda,[0]}) \notag \\
  &\qquad \qquad- \alpha_{\psi, -}^{-1} i \diamond_\Xi (\overline{U_f}^\dagger \Sigma^\lambda \gamma^0 \overline{U_f}, \alpha_{\psi, +}^{-1}) -\frac{i}{2} \big\{ \Xi, \partial_k \gamma^0 \hat{S}^{\lambda, [0]} \big\}.
  \label{eq:sol_masseigenstate_fer}
\end{align}
We have defined the covariant diamond operator
\begin{align}
  \diamond_\Xi (A ,B) &\equiv \frac{1}{2} \Big( (\partial_x A - [\Xi, A]) \partial_{k} B - \partial_{k} A (\partial_x B - [\Xi, B]) \Big) \notag \\
  &= \diamond (A,B) -\frac{1}{2} \Big( [\Xi, A] \partial_{k} B - \partial_k A [\Xi, B] \Big),
\end{align}
where the connection $\Xi$ is defined by
\begin{align}
  \Xi = \begin{cases}
    (\partial_x U_b) U_b^\dagger~~~~~(\text{bosons}) \\
    (\partial_x \overline{U_f}^\dagger) \overline{U_f}~~~(\text{fermions}) \\
  \end{cases}.
\end{align}

\section{Derivation of the Boltzmann equation \label{sec:systematic}}

In this section, we derive the Boltzmann equation for the quasiparticle by applying the prescriptions introduced in section~\ref{sec:approx} to the solutions derived in section~\ref{sec:solution}.
We first show that the Breit--Wigner-type structure in the spectral function reduces to delta-function contributions under the quasiparticle approximation, both at zeroth order and at first order in the derivative expansion.
We then identify the on-shell part of the Wightman functions.
By substituting the on-shell part into the Hermitian equation in Eq.~\eqref{eq:KBeq_1}, we derive the Boltzmann equation.

\subsection{Spectral function with the quasiparticle approximation \label{sec:spectral}}

In this subsection, we apply the quasiparticle approximation to the spectral function obtained up to first order in the derivative expansion.
We show that, after taking the NWL, the spectral function can be written in terms of delta functions imposing the on-shell condition.
For fermions, the correction at first order in the derivative expansion shifts the quasiparticle shell, which is consistent with the conventional semiclassical approach.

\subsubsection{Bosons ($n=0$)}

We first consider the bosonic case at zeroth order in the derivative expansion.
The self-energy combination $U_b \Pi U_b^\dagger$ entering $\alpha_{\phi,\pm}$ is, in general, non-diagonal in flavor space.
Based on the discussion in section~\ref{sec:mixing}, the self-energy can be decomposed into the flavor-universal part and the flavor-dependent remainder (see Eq.~\eqref{eq:flavorperturbation}).
After the local-basis rotation, the former remains diagonal in the local mass basis, while the latter is, in general, non-diagonal and assumed to be small:
\begin{align}
  (U_b \Pi U_b^\dagger)_{ij}
  =
  \overline{\Pi} \, \delta_{ij}
  +
  (U_b  \delta \Pi  U_b^\dagger)_{ij}.
  \label{eq:boson_selfenergy_decompose}
\end{align}
By substituting Eq.~\eqref{eq:boson_selfenergy_decompose} into Eq.~\eqref{eq:def_alpha_bos}, we have 
\begin{align}
  \alpha_{\phi, \pm}
  =
  \sum_i
  (\Omega_i^2 \mp i \overline{\Pi}_{\mathcal{A}})P_i
  +
  \left(
  - U_b \delta \Pi_{\mathcal{H}} U_b^\dagger
  \mp i U_b \delta \Pi_{\mathcal{A}} U_b^\dagger
  \right),
\end{align}
where 
\begin{align}
  \Omega_i^2 = k^2 - (m^2_i + \overline{\Pi}_{\mathcal{H}}),
\end{align}
and $P_i$ is the projection matrix onto the $i$-th flavor mode, which satisfies $P_i P_j = \delta_{ij} P_i$ and $\sum_i P_i = 1$.
Applying the basis transformation given in Eq.~\eqref{eq:masstoflavor} to Eq.~\eqref{eq:sol_n0} and expanding the inverse of $\alpha_{\phi, \pm}$ to first order in the flavor-dependent part, we obtain the advanced and retarded propagators
\begin{align}
  \hat{\Delta}^{a,r,[0]} &= \alpha_{\phi, \pm}^{-1} \notag \\
  &= \sum_i (\Omega_i^2 \mp i \overline{\Pi}_{\mathcal{A}})^{-1} P_i \notag \\
  &- \sum_{i,j}
  (\Omega_i^2 \mp i \overline{\Pi}_{\mathcal{A}})^{-1}
  (\Omega_j^2 \mp i \overline{\Pi}_{\mathcal{A}})^{-1}
  P_i
  \left(
  - U_b \delta \Pi_{\mathcal{H}} U_b^\dagger
  \mp i U_b \delta \Pi_{\mathcal{A}} U_b^\dagger
  \right)
  P_j
  + \cdots.
  \label{eq:bos_n0_ra}
\end{align}
Neglecting the flavor-dependent self-energy $\delta \Pi_{\mathcal{H},\mathcal{A}}$ in $\alpha^{-1}_{\phi, \pm}$ and substituting them into Eq.~\eqref{eq:sol_masseigenstate_bos}, we obtain the spectral function 
\begin{align}
  \hat{\mathscr{A}}_\phi^{[0]} = \sum_i P_i \frac{\overline{\Pi}_{\mathcal{A}}}{\Omega_i^4 + \overline{\Pi}_{\mathcal{A}}^2}.
  \label{eq:boson_spectral0}
\end{align}
In the quasiparticle approximation, the spectral function in each flavor mode takes the on-shell form,
\begin{align}
  P_i \hat{\mathscr{A}}^{[0]}_\phi
  &\to
  P_i \, \pi  \mathrm{sign}(k^0) \, \delta(\Omega_i^2) .
  \label{eq:spectral_boson_qpa}
\end{align}

Similarly, using Eq.~\eqref{eq:bos_n0_ra} within the flavor-universal approximation and the definition of the Hermitian part in Eq.~\eqref{eq:propag3}, we obtain
\begin{align}
  P_i \hat{\Delta}_{\mathcal{H}}^{[0]}
  = P_i \frac{\Omega^2_i}{\Omega_i^4 + \overline{\Pi}_{\mathcal{A}}^2},
\end{align}
which reduces to an off-shell principal-value (PV) contribution in the quasiparticle approximation,
\begin{align}
  P_i \hat{\Delta}_{\mathcal{H}}^{[0]}
  \to
  P_i \times \mathrm{PV}\left(\frac{1}{\Omega_i^2}\right) .
  \label{eq:bos_hermitian_propag}
\end{align}
Therefore, the Hermitian part does not contain an on-shell part in the NWL.

\subsubsection{Bosons ($n=1$)} 

Even in the flavor-universal approximation in the self-energy, the solutions in the local mass basis are not necessarily closed within the flavor-diagonal sector.
This is because the spacetime dependence of the local basis transformation generates connection terms.
As seen in Eq.~\eqref{eq:sol_masseigenstate_bos}, these terms appear not only through the Wigner transformation, such as $-\frac{i}{2}\{\Xi,\partial_k\hat{\mathscr{A}}_\phi^{[0]}\}$, but also through the replacement of the diamond operator by the covariant one, $\diamond \to \diamond_\Xi$.

As discussed in section~\ref{sec:basis}, the diagonal element of $\Xi$ can be removed by a local phase transformation.
We choose the basis in which $\Xi_{ii}=0$.
In this convention, since the zeroth-order spectral function is diagonal in flavor space, the connection terms do not generate a diagonal correction directly.
On the other hand, the off-diagonal connection $\Xi_{ij}$ can generally generate off-diagonal components of the spectral function at first order in the derivative expansion.

In the following discussions, we specify the on-shell part of the spectral and Wightman functions by retaining only the flavor-diagonal elements in the local mass basis.
The quasiparticle shells used below are therefore determined by the diagonal components in the local mass basis.
This treatment is equivalent to the diagonal limit used in refs.~\cite{Kainulainen:2001cn,Prokopec:2003pj}\footnote{
The off-diagonal components induced by the connection terms describe quantum coherence between different flavor modes in the local mass basis.
As discussed in ref.~\cite{Prokopec:2003pj}, their feedback to the flavor-diagonal equations does not give an independent on-shell contribution at first order in the derivative expansion, provided the mass eigenvalues are not nearly degenerate.
These off-diagonal components were treated explicitly in refs.~\cite{Konstandin:2004gy,Konstandin:2005cd,Herranen:2008hi,Cirigliano:2009yt,Cirigliano:2011di} for flavor mixing with a spacetime-dependent complex mass matrix.
In the present work, by contrast, we focus on the flavor-diagonal part in our finite-width formulation.
Extending our construction to include these off-diagonal components would require keeping the full flavor matrices of the propagators and self-energy.
We leave this extension for future work.
}.

As can be seen in Eq.~\eqref{eq:sol_masseigenstate_bos}, the contributions at first order in the derivative expansion involve $\diamond_\Xi (\alpha_{\phi, \pm}, \alpha_{\phi, \pm}^{-1})$.
However, using the flavor-universal approximation for the self-energy, we have
\begin{align}
  \diamond (\alpha_{\phi, \pm}, \alpha_{\phi, \pm}^{-1}) = 0.
  \label{eq:bosonic_zerocontribution}
\end{align}
As a result, in the basis in which $\Xi_{ii}=0$, the remaining terms at first order are induced by the connection terms and generate only off-diagonal components.
In the flavor-diagonal limit adopted here, these off-diagonal components are not retained.
Using Eq.~\eqref{eq:bosonic_zerocontribution} in Eqs.~\eqref{eq:sol_n1} and \eqref{eq:sol_masseigenstate_bos}, together with the definition in Eq.~\eqref{eq:propag3}, we obtain
\begin{align}
  \big(\hat{\Delta}^{r,a,[1]}\big)_{ii} = 0,~~~~
  \big(\hat{\mathscr{A}}^{[1]}_\phi \big)_{ii} = 0, ~~~~ \big(\hat{\Delta}_{\mathcal{H}}^{[1]} \big)_{ii} = 0.
  \label{eq:boson_spectral1}
\end{align}
Therefore, the quasiparticle shell is not shifted at first order in the derivative expansion for the bosonic system under the present setup\footnote{
    At second order in the derivative expansion, scalar fields are known to acquire a CP-even correction to the shell function~\cite{Joyce:1999fw,Joyce:1999uf}.
}.

\subsubsection{Fermions ($n=0$)}

We discuss the fermionic case in a similar way to the bosonic case.
Based on the spinor decomposition in Eq.~\eqref{eq:spinordecompose} and the flavor-universal approximation discussed in section~\ref{sec:mixing}, the flavor-universal part of the self-energy can be decomposed into (spin) $\otimes$ (chiral) $\otimes$ (flavor) spaces.
We denote the decomposed self-energy functions as
\begin{align}
  \overline{U}^\dagger \Sigma \gamma^0 \overline{U}
  =
  \sum_{i,s} P_s \otimes
  \Big( \overline{a}_{0s} + \sum_m \overline{a}_{ms}\sigma^m \Big)
  \otimes P_i
  + \overline{U}^\dagger \delta\Sigma\gamma^0\overline{U},
  \label{eq:sigmadecompose}
\end{align}
where the first term on the right-hand side is the flavor-universal part and the second term is the flavor-dependent remainder.
Within the flavor-universal approximation, we neglect the second term.
Substituting Eq.~\eqref{eq:sigmadecompose} into Eq.~\eqref{eq:def_alpha_fer}, in the boosted frame, we have
\begin{align}
  P_s P_i \alpha_{\psi, \pm} &=  P_s P_i \big( \overline{\mathcal{N}}_{s,i}^{[0]} \pm i \overline{\mathcal{N}}_{\mathcal{A}}^{[0]} \big) ,
  \label{eq:alpha_altanative}
\end{align}
where 
\begin{align}
  \overline{\mathcal{N}}_{s,i}^{[0]} &\equiv \tilde{\mathcal{K}}^0  - \mathcal{M}_{Ri} \sigma^1 - \mathcal{M}_{Ii} \sigma^2 - \mathcal{K}_z \sigma^3, \notag \\
  \overline{\mathcal{N}}_{\mathcal{A}}^{[0]} &\equiv - \overline{a}_{0s,\mathcal{A}} -\overline{a}_{1s,\mathcal{A}} \sigma^1 -\overline{a}_{2s,\mathcal{A}}\sigma^2 - \overline{a}_{3s,\mathcal{A}}\sigma^3,
\end{align}
and 
\begin{align}
  \tilde{\mathcal{K}}^0 \equiv \tilde{k}^0 - \overline{a}_{0s,\mathcal{H}}, ~~
  \mathcal{M}_{Ri} \equiv m_{Ri} + \overline{a}_{1s,\mathcal{H}} , ~~
  \mathcal{M}_{Ii} \equiv m_{Ii} + \overline{a}_{2s,\mathcal{H}} ,~~
  \mathcal{K}_z \equiv s k_z + \overline{a}_{3s,\mathcal{H}} .
  \label{eq:def_coef_fer}
\end{align}
Applying the basis transformation in Eq.~\eqref{eq:masstoflavor} to Eq.~\eqref{eq:sol_n0} and substituting Eq.~\eqref{eq:alpha_altanative} into the resulting expression, we obtain the retarded and advanced propagators as
\begin{align}
  P_s P_i \gamma^0 \hat{S}^{a,r,[0]} = P_s P_i \alpha_{\psi, \pm}^{-1} = P_s P_i \frac{\mathcal{N}_{s,i}^{[0]} \pm i \mathcal{N}_{\mathcal{A}}^{[0]} }{\Omega_{s,i}^2 \mp i \Gamma_{s,i}},
  \label{eq:fermion_n0_ra}
\end{align}
where 
\begin{align}
  \Omega_{s,i}^2 &= (\tilde{\mathcal{K}}^0)^2 - (\mathcal{K}_z)^2 - (\mathcal{M}_{Ri})^2 - (\mathcal{M}_{Ii})^2 + \mathcal{O}(\overline{a}_{\mathcal{A}}^2), \notag \\
  \Gamma_{s,i} &= 2 \Big( \tilde{\mathcal{K}}^0 \overline{a}_{0s,\mathcal{A}} + \mathcal{K}_z \overline{a}_{3s,\mathcal{A}} + \mathcal{M}_{Ri} \overline{a}_{1s,\mathcal{A}} + \mathcal{M}_{Ii} \overline{a}_{2s,\mathcal{A}} \Big), \notag \\
  \mathcal{N}_{s,i}^{[0]} &= \tilde{\mathcal{K}}^0  + \mathcal{M}_{Ri} \sigma^1 + \mathcal{M}_{Ii} \sigma^2 + \mathcal{K}_z \sigma^3, \notag \\
  \mathcal{N}_{\mathcal{A}}^{[0]} &= - \overline{a}_{0s,\mathcal{A}} + \overline{a}_{1s,\mathcal{A}} \sigma^1 + \overline{a}_{2s,\mathcal{A}}\sigma^2 + \overline{a}_{3s,\mathcal{A}}\sigma^3.
  \label{eq:fermion_omega_gamma}
\end{align}
We note that the terms of order $\overline{a}_{\mathcal{A}}^2$ in $\Omega_{s,i}^2$ can be neglected, since they do not affect the quasiparticle shell in the NWL.
As a result, by substituting $\alpha_{\psi, \pm}^{-1}$ into the spectral function in Eq.~\eqref{eq:sol_masseigenstate_fer}, we obtain
\begin{align}
  P_s P_i \gamma^0 \hat{\mathscr{A}}^{[0]}_\psi &= P_s P_i \frac{\Gamma_{s,i} \mathcal{N}_{s,i}^{[0]}}{\Omega_{s,i}^4 + \Gamma_{s,i}^2} + P_s P_i \frac{\Omega_{s,i}^2 \mathcal{N}_{\mathcal{A}}^{[0]} }{\Omega_{s,i}^4 + \Gamma_{s,i}^2} \notag \\
  &\to P_s P_i \, \pi \mathrm{sign}(\tilde{k}^0) \, \delta(\Omega_{s,i}^2) \mathcal{N}_{s,i}^{[0]},
  \label{eq:fermion_spectral0}
\end{align}
where the quasiparticle approximation is taken in the second line.
Since $\mathcal{N}_{\mathcal{A}}^{[0]}=\mathcal{O}(\overline{a}_{0,\mathcal{A}})$ in the NWL, the second term in the first line vanishes and therefore does not contribute to the on-shell part.

Using Eq.~\eqref{eq:fermion_n0_ra} and the definition of the Hermitian part in Eq.~\eqref{eq:propag3}, we obtain
\begin{align}
  P_s P_i \gamma^0 \hat{S}_{\mathcal{H}}^{[0]} &= P_s P_i\frac{\Omega_{s,i}^2 \mathcal{N}^{[0]}_{s,i}}{\Omega_{s,i}^4 + \Gamma_{s,i}^2} - P_s P_i\frac{\Gamma_{s,i} \mathcal{N}_{\mathcal{A}}^{[0]}}{\Omega_{s,i}^4 + \Gamma_{s,i}^2} \notag \\
  &\to P_s P_i \times \mathrm{PV}\left(\frac{1}{\Omega_{s,i}^2}\right)
  \mathcal{N}^{[0]}_{s,i}
  \label{eq:fer_hermitian_propag}
\end{align}
where the quasiparticle approximation is taken in the second line.
Therefore, as in the bosonic case, the Hermitian part reduces to an off-shell principal-value contribution in the quasiparticle approximation.

\subsubsection{Fermions ($n=1$)}

As in the bosonic case, there are contributions from the connection terms.
In the basis with $\Xi_{ii}=0$, the mass term is given by Eq.~\eqref{eq:mass_replace}, while its absolute value remains unchanged.
For the derivative of its phase, we make the replacement $\theta_i^\prime \to \theta_i^\prime+\Delta^{LR}_i$, where
\begin{align}
  \Delta_i^{LR} = \big[ i (\partial_z U_L)U_L^\dagger - i (\partial_z U_R)U_R^\dagger \big]_{ii},
\end{align}
so that the flavor-diagonal elements in Eq.~\eqref{eq:fermion_connection_trans} vanish.
In the following, $m_{Ri}$ and $m_{Ii}$ denote the real and imaginary parts of the complex mass in this basis, respectively.
Although the off-diagonal connection terms still remain, as discussed in the bosonic case, we focus only on the flavor-diagonal elements of the spectral and Wightman functions.

After transforming Eq.~\eqref{eq:sol_n1} to the mass basis using Eq.~\eqref{eq:masstoflavor} and substituting Eq.~\eqref{eq:alpha_altanative}, the retarded and advanced propagators at first order in the derivative expansion take the form
\begin{align}
  P_s P_i \gamma^0 \hat{S}^{a,r,[1]} &= P_s P_i \alpha_{\psi, \pm}^{-1} i \diamond (\alpha_{\psi, \pm}, \alpha_{\psi, \pm}^{-1}) = P_s P_i \frac{\mathcal{N}_{s,i}^{[1]} \pm i \mathcal{N}_{\mathcal{A}}^{[1]}}{(\Omega^2_{s,i} \mp i \Gamma_{s,i})^2},
  \label{eq:fer_n1_ra}
\end{align}
where 
\begin{align}
  \mathcal{N}_{s,i}^{[1]} &= \frac{-i}{2} \bigg[ \big( \partial_z \mathcal{N}_{s,i}^{[0]} \big) \big(\partial_{k_z} \overline{\mathcal{N}}_{s,i}^{[0]} \big) \mathcal{N}_{s,i}^{[0]} - \mathcal{N}_{s,i}^{[0]} \big( \partial_{k_z} \overline{\mathcal{N}}_{s,i}^{[0]} \big) \big( \partial_z \mathcal{N}_{s,i}^{[0]} \big) + \mathcal{O}(\overline{a}_{\mathcal{A}}^2) \bigg] \notag \\
  &= \bigg[m_{Ii}^\prime(\mathcal{K}_z \partial_{k_z} \mathcal{M}_{Ri} -  \mathcal{M}_{Ri} \partial_{k_z} \mathcal{K}_z ) - m_{Ri}^\prime(\mathcal{K}_z \partial_{k_z} \mathcal{M}_{Ii} -  \mathcal{M}_{Ii} \partial_{k_z} \mathcal{K}_z ) \notag \\
  &\qquad + m_{Ii}^\prime (\mathcal{K}_z \partial_{k_z} \tilde{\mathcal{K}}^0 - \tilde{\mathcal{K}}^0 \partial_{k_z} \mathcal{K}_z) \sigma^1 - m_{Ri}^\prime (\mathcal{K}_z \partial_{k_z} \tilde{\mathcal{K}}^0 - \tilde{\mathcal{K}}^0 \partial_{k_z} \mathcal{K}_z) \sigma^2 \notag \\
  &\qquad +\Big[ m_{Ii}^\prime (\tilde{\mathcal{K}}^0 \partial_{k_z}\mathcal{M}_{Ri} - \mathcal{M}_{Ri} \partial_{k_z}\tilde{\mathcal{K}}^0) - m_{Ri}^\prime (\tilde{\mathcal{K}}^0 \partial_{k_z}\mathcal{M}_{Ii} - \mathcal{M}_{Ii} \partial_{k_z}\tilde{\mathcal{K}}^0) \Big]\sigma^3   + \mathcal{O}(\overline{a}_{\mathcal{A}}^2) \bigg], \notag \\
  \mathcal{N}_{\mathcal{A}}^{[1]} &= \frac{-i}{2} \bigg[ \big( \partial_z \mathcal{N}_{s,i}^{[0]} \big) \big(\partial_{k_z} \overline{\mathcal{N}}_{\mathcal{A}}^{[0]} \big) \mathcal{N}_{s,i}^{[0]} - \mathcal{N}_{s,i}^{[0]} \big( \partial_{k_z} \overline{\mathcal{N}}_{\mathcal{A}}^{[0]} \big) \big( \partial_z \mathcal{N}_{s,i}^{[0]} \big) \notag \\
  &\qquad + \big( \partial_z \mathcal{N}_{s,i}^{[0]} \big) \big(\partial_{k_z} \overline{\mathcal{N}}_{s,i}^{[0]} \big) \mathcal{N}_{\mathcal{A}}^{[0]} - \mathcal{N}_{\mathcal{A}}^{[0]} \big( \partial_{k_z} \overline{\mathcal{N}}_{s,i}^{[0]} \big) \big( \partial_z \mathcal{N}_{s,i}^{[0]} \big) \bigg].
\end{align}

Substituting the expressions in Eqs.~\eqref{eq:alpha_altanative} and \eqref{eq:fermion_n0_ra} into the first-order spectral function in Eq.~\eqref{eq:sol_masseigenstate_fer}, we obtain
\begin{align}
  P_s P_i \gamma^0 \hat{\mathscr{A}}^{[1]}_{\psi} &= P_s P_i \frac{2 \Gamma_{s,i} \Omega^2_{s,i}}{(\Omega_{s,i}^4 + \Gamma_{s,i}^2)^2} \mathcal{N}_{s,i}^{[1]} + P_s P_i \frac{\Omega^4_{s,i} - \Gamma_{s,i}^2}{(\Omega_{s,i}^4 + \Gamma_{s,i}^2)^2} \mathcal{N}_{\mathcal{A}}^{[1]} \notag \\
  &\to - P_s P_i \, \pi \mathrm{sign}(\tilde{k}^0) \delta^\prime \big( \Omega_{s,i}^2 \big) \mathcal{N}_{s,i}^{[1]},
  \label{eq:fermion_spectral1}
\end{align}
where the quasiparticle approximation is taken in the second line.
Since $\mathcal{N}_{\mathcal{A}}^{[1]}=\mathcal{O}(\overline{a}_{0, \mathcal{A}})$ in the NWL, the last term in the first line again vanishes.
The derivative of the delta function indicates that the correction at first order shifts the quasiparticle shell for each component in chiral space.
At this order, this shell shift follows from the expansion, $\delta(x+\delta x)  = \delta(x)+\delta x \delta^\prime(x) +\mathcal{O}(\delta x^2)$.

Using Eq.~\eqref{eq:fer_n1_ra} and the definition of the Hermitian part in Eq.~\eqref{eq:propag3}, we obtain
\begin{align}
  P_s P_i \gamma^0 \hat{S}_{\mathcal{H}}^{[1]} &= P_s P_i \frac{\Omega^4_{s,i} - \Gamma_{s,i}^2}{(\Omega_{s,i}^4 + \Gamma_{s,i}^2)^2} \mathcal{N}_{s,i}^{[1]} - P_s P_i \frac{2 \Omega_{s,i}^2 \Gamma_{s,i}}{(\Omega_{s,i}^4 + \Gamma_{s,i}^2)^2} \mathcal{N}_{\mathcal{A}}^{[1]} \notag \\
  &\to P_s P_i \times \mathrm{PV} \Big(\frac{1}{\Omega_{s,i}^4} \Big) \mathcal{N}_{s,i}^{[1]},
  \label{eq:hermitian_propag_fer_n1}
\end{align}
where the off-shell principal-value distribution is obtained in the second line in the quasiparticle approximation.

\subsection{On-shell part of the Wightman functions \label{sec:Wightman}}

In this subsection, we define the flavor-diagonal on-shell part of the Wightman functions up to first order in the derivative expansion, which is used in deriving the Boltzmann equation.
As discussed in section~\ref{sec:selfenergydecompose}, the on-shell part of the Wightman functions is extracted by substituting $g^\lambda = g^\lambda_{\mathrm{stat}}$ into Eqs.~\eqref{eq:sol_masseigenstate_bos}-\eqref{eq:sol_masseigenstate_fer}.

\subsubsection{Bosons}

Up to first order in the derivative expansion, the bosonic Wightman functions can be written as
\begin{align}
  \hat{\Delta}^\lambda
  =
  \hat{\Delta}^{\lambda,[0]} + \hat{\Delta}^{\lambda,[1]} .
\end{align}
By substituting $\Pi^\lambda = \Pi^\lambda_{\mathrm{stat}}$ into the solution given in Eq.~\eqref{eq:sol_masseigenstate_bos}, in the flavor-universal approximation, we obtain
\begin{align}
  P_i \hat{\Delta}^{\lambda}  &= \frac{2}{i} n_i^\lambda P_i \hat{\mathscr{A}}_\phi^{[0]} + P_i \, \frac{i}{2} \frac{4 \overline{\Pi}_{\mathcal{A}}^2 }{\big( \Omega_i^4 + \overline{\Pi}_{\mathcal{A}}^2 \big)^2} \big\{ \Omega_i^2 , n_i^\lambda \big\}_{z,k_z},
  \label{eq:boson_Wightman_shell_before}
\end{align}
where we have introduced the Poisson bracket
\begin{align}
  \big\{X, Y \big\}_{z,k_z} \equiv \partial_z X \, \partial_{k_z} Y - \partial_{k_z} X \, \partial_{z} Y.
\end{align}
The subscript on $n^\lambda_i$ explicitly denotes the statistical function for the $i$-th flavor component.
Here we have neglected the $\partial_{k_z}\overline{\Pi}_{\mathcal{A}}$ terms for simplicity; keeping or neglecting these terms does not affect the following derivation of the Boltzmann equation.

The first term on the right-hand side of Eq.~\eqref{eq:boson_Wightman_shell_before} is proportional to the spectral function at zeroth order in the derivative expansion.
In the quasiparticle approximation, this term reduces to the delta-function contribution as shown in Eq.~\eqref{eq:spectral_boson_qpa}.
Since the spectral function receives no first-order correction in the present approximation, the quasiparticle shell is not shifted by the gradient of the mass term.

We do not identify the second term on the right-hand side of Eq.~\eqref{eq:boson_Wightman_shell_before} as part of the on-shell Wightman function.
This term is associated with the Liouville term for the statistical function $n_i^\lambda$ (cf. Eq.~\eqref{eq:boson_liouville}).
It appears because the Kadanoff--Baym equation for the Wightman functions in Eq.~\eqref{eq:Wightman2}, which is solved order by order in the derivative expansion in section~\ref{sec:solution}, contains information from both the constraint and kinetic equations.
It therefore describes the kinetic evolution of $n_i^\lambda$, rather than a correction to the quasiparticle shell, and is not included in the on-shell Wightman function used to derive the Boltzmann equation\footnote{
Away from the quasiparticle shell, where $\Omega_i^2 \neq 0$, the second term of the right-hand side in Eq.~\eqref{eq:boson_Wightman_shell_before} vanishes in the NWL.
Near the quasiparticle shell, however, it may appear to be singular as $1/\overline{\Pi}_{\mathcal{A}}$.
This apparent singularity at first order in the derivative expansion is removed once $n_i^\lambda$ is constrained by the kinetic equation at the same order.
Following the procedure described in section~\ref{sec:Boltzmann}, substituting the zeroth-order on-shell Wightman functions,
$P_i \hat{\Delta}^{\lambda,[0]} = \frac{2}{i} n_i^\lambda P_i \hat{\mathscr{A}}_\phi^{[0]}$,
into the Hermitian equation in Eq.~\eqref{eq:KBeq_1} gives
\begin{align}
  2 \pi \mathrm{sign}(k^0) \delta(\Omega_i^2) \big\{ \Omega_i^2 , n_i^\lambda \big\}_{z,k_z}
  = C_i.
  \label{eq:boson_Boltzmann_balance}
\end{align}
The collision term on the right-hand side is of the same order in the interaction strength as $\overline{\Pi}_{\mathcal{A}}$.
Upon imposing Eq.~\eqref{eq:boson_Boltzmann_balance} on $n_i^\lambda$, the second term in Eq.~\eqref{eq:boson_Wightman_shell_before} becomes proportional to $C_i/\overline{\Pi}_{\mathcal{A}}$ near the quasiparticle shell and therefore remains finite in the weak-coupling limit.
}.

Accordingly, retaining only the first term in Eq.~\eqref{eq:boson_Wightman_shell_before}, we specify the on-shell part of the Wightman functions in the quasiparticle approximation as
\begin{align}
  P_i i \hat{\Delta}^{\lambda}_{\mathrm{shell}} = P_i \times 2 \pi n_i^\lambda \mathrm{sign}(k^0) \delta(\Omega_i^2) .
  \label{eq:boson_Wightman_shell}
\end{align}

\subsubsection{Fermions}

Up to first order in the derivative expansion, the fermionic Wightman functions can be written as
\begin{align}
  \hat{S}^\lambda
  =
  \hat{S}^{\lambda,[0]} + \hat{S}^{\lambda,[1]}.
\end{align}
By substituting $\Sigma^\lambda = \Sigma^\lambda_{\mathrm{stat}}$ into Eq.~\eqref{eq:sol_masseigenstate_fer}, we obtain
\begin{align}
  P_s P_i (-i \gamma^0 \hat{S}^{\lambda}) &= 2 n^\lambda_{s, i} P_s P_i \big( \gamma^0 \hat{\mathscr{A}}^{[0]}_\psi + \gamma^0 \hat{\mathscr{A}}^{[1]}_\psi \big) \notag \\
  &+ P_s P_i \, \frac{i}{\Omega_{s,i}^4 + \Gamma_{s,i}^2} \bigg( \Big\{ \mathcal{N}_{s,i}^{[0]}, n_{s,i}^\lambda \Big\}_{z,k_z} \overline{\mathcal{N}}_{s,i}^{[0]} \mathcal{N}_{\mathcal{A}}^{[0]}   - \mathcal{N}_{\mathcal{A}}^{[0]} \overline{\mathcal{N}}_{s,i}^{[0]}  \Big\{ \mathcal{N}_{s,i}^{[0]}, n_{s,i}^\lambda \Big\}_{z,k_z}  \bigg) \notag \\
  &- P_s P_i \, \frac{1}{2} \frac{4 \Gamma_{s,i}^2 }{\big( \Omega_{s,i}^4 + \Gamma_{s,i}^2 \big)^2} \big\{ \Omega_{s,i}^2 , n_{s,i}^\lambda \big\}_{z,k_z} \mathcal{N}_{s,i}^{[0]} + (\text{vanishing terms in the NWL}).
  \label{eq:fermion_Wightman_QPA}
\end{align}
Again, terms containing $\partial_{k_z}\bar a_{\mathcal{A}}$ are not displayed explicitly.
These terms do not give additional contributions in the quasiparticle approximation.
The spin and flavor indices of $n^\lambda_{s,i}$ are shown explicitly.

The first term on the right-hand side in Eq.~\eqref{eq:fermion_Wightman_QPA} shows that, in the quasiparticle approximation, the first-order correction proportional to $\mathcal{N}^{[1]}_{s,i}\delta^\prime(\Omega_{s,i}^2)$ shifts the quasiparticle shell.
In the quasiparticle approximation, the second term becomes 
\begin{align}
  &\frac{i}{\Omega_{s,i}^4 + \Gamma_{s,i}^2} \bigg( \Big\{ \mathcal{N}_{s,i}^{[0]}, n_{s,i}^\lambda \Big\}_{z,k_z} \overline{\mathcal{N}}_{s,i}^{[0]} \mathcal{N}_{\mathcal{A}}^{[0]}   - \mathcal{N}_{\mathcal{A}}^{[0]} \overline{\mathcal{N}}_{s,i}^{[0]}  \Big\{ \mathcal{N}_{s,i}^{[0]}, n_{s,i}^\lambda \Big\}_{z,k_z}  \bigg) \notag \\
  &\to -i \frac{1}{2 \tilde{\mathcal{K}}^0} \pi \mathrm{sign}(\tilde{k}^0) \delta \big( \Omega_{s,i}^2 \big) \bigg[ \Big\{ \mathcal{N}_{s,i}^{[0]}, n_{s,i}^\lambda \Big\}_{z,k_z}, \, \overline{\mathcal{N}}_{s,i}^{[0]}  \bigg].
\end{align}
Therefore, it also contributes to the on-shell part of the Wightman functions.
As in the bosonic case, the third term in Eq.~\eqref{eq:fermion_Wightman_QPA} is associated with the Liouville term for the statistical function $n^\lambda_{s,i}$.
It therefore describes the kinetic evolution of $n^\lambda_{s,i}$, rather than a correction to the quasiparticle shell, and is not included in the on-shell part of the Wightman functions\footnote{
    As in the bosonic case, this term may appear to generate a contribution proportional to $1/\overline{a}_{0,\mathcal{A}}$ in the NWL.
    However, this apparent singularity is removed once the zeroth-order on-shell Wightman functions are constrained by the Hermitian equation at first order, and the resulting contribution remains finite in the weak-coupling limit.
}.
The on-shell part is then defined as
\begin{align}
  P_s P_i(-i \gamma^0 \hat{S}^{\lambda}_{\text{shell}}) =  P_s P_i \times 2 \pi \mathrm{sign}(\tilde{k}^0) &\Bigg( n^\lambda_{s,i} \delta \big( \Omega_{s,i}^2 \big) \mathcal{N}^{[0]}_{s,i} - n^\lambda_{s,i} \delta^\prime \big( \Omega_{s,i}^2 \big) \mathcal{N}^{[1]}_{s,i} \notag \\
  &\qquad -i \frac{1}{4 \tilde{\mathcal{K}}^0} \delta \big( \Omega_{s,i}^2 \big) \bigg[ \Big\{ \mathcal{N}_{s,i}^{[0]}, n_{s,i}^\lambda \Big\}_{z,k_z}, \, \overline{\mathcal{N}}_{s,i}^{[0]}  \bigg] \Bigg).
  \label{eq:fermion_shell_n1}
\end{align}
We note that the on-shell Wightman functions obtained here reduce to those derived in refs.~\cite{Kainulainen:2001cn,Prokopec:2003pj,Kainulainen:2002th} in the limit where the self-energy corrections are neglected.

\subsection{Deriving the Boltzmann equation \label{sec:Boltzmann}}

Finally, within the present approximations, we derive the Boltzmann equation governing the kinetic evolution of the quasiparticle distribution.

The term involving $\mathcal{D}_K$ in the Hermitian equation in Eq.~\eqref{eq:KBeq_1} is the derivative term associated with the kinetic evolution of the Wightman functions.
Since it contains one derivative, applying it to the first-order correction to the on-shell Wightman functions gives a second-order contribution in the derivative expansion.
Accordingly, by expanding $e^{-i \diamond}$ up to second order in Eq.~\eqref{eq:KBeq_1} and taking the trace, we obtain\footnote{
This trace form is used only after inserting the on-shell solutions.
For example, $\mathrm{Tr}[P_i X]$ selects the diagonal component associated with the quasiparticle of the $i$-th flavor, rather than summing over the Boltzmann equations for all flavors.
The off-diagonal Wightman functions are not kept in the present description.
}
\begin{align}
  H[G^\lambda] = C[G^\lambda],
  \label{eq:schematical_boltzmann}
\end{align}
where 
\begin{align}
  H[G^\lambda] &\equiv -2i \mathrm{Tr}\big[ \diamond \big(\mathcal{D}_C - \mathcal{M} - g_{\mathcal{H}}, G^\lambda \big) \big], \notag \\
  C[G^\lambda] &\equiv
  \mathrm{Tr} \big[ g^> G^< - g^< G^> \big] - \frac{1}{2}
  \mathrm{Tr} \big[\diamond^2
  \big( g^>, G^< \big) - \diamond^2 \big( g^<,  G^> \big) \big].
  \label{eq:hermitian_trace_for_boltzmann}
\end{align}
Here we have used the cyclicity of the trace and $\mathrm{Tr}[\diamond(A,B)] = -\mathrm{Tr}[\diamond(B,A)]$.
In addition, we have dropped the term involving $G_{\mathcal H}$ on the left-hand side of Eq.~\eqref{eq:KBeq_1}, since the Hermitian propagator does not contain an on-shell part in the quasiparticle approximation, as illustrated in Eqs.~\eqref{eq:bos_hermitian_propag}, \eqref{eq:fer_hermitian_propag}, and \eqref{eq:hermitian_propag_fer_n1}.
This term therefore does not contribute to the Boltzmann equation derived below.

To derive the Boltzmann equation for the quasiparticles, we evaluate these functionals in the local mass basis and substitute the on-shell part of the Wightman functions defined in that basis.
By taking the projection with $\mathcal{P}=P_i$ and $\mathcal{P}=P_sP_i$ for bosons and fermions, respectively, and integrating Eq.~\eqref{eq:schematical_boltzmann} over $k^0 > 0$, we have
\begin{align}
  \int_0^\infty \frac{d k^0}{2\pi}
  H[\mathcal{P}\,\hat{G}^\lambda_{\mathrm{shell}}]
  =
  \int_0^\infty \frac{d k^0}{2\pi} C[\mathcal{P}\,\hat{G}^\lambda_{\mathrm{shell}}].
  \label{eq:Liouville_vs_collision}
\end{align}
The left-hand side corresponds to the Liouville term evaluated in the local mass basis, while the right-hand side is the collision term evaluated with the on-shell part of $\hat{G}^\lambda$ and the Wightman self-energy $\hat{g}^\lambda$, both expressed in this basis.

The $\diamond^2$ term in Eq.~\eqref{eq:hermitian_trace_for_boltzmann} does not contribute to the Boltzmann equation at second order in the derivative expansion; the contribution involving $\delta\hat g_{\mathrm{coll}}$ in this term is of third order or higher in the derivative expansion, while the contribution involving $\hat g^\lambda_{\mathrm{stat}}\propto n^\lambda \hat g_{\mathcal A}$ vanishes in the NWL and therefore does not give an additional on-shell contribution.
Consequently, up to second order in the derivative expansion, the collision term becomes
\begin{align}
  \int_0^\infty \frac{d k^0}{2\pi} C[\mathcal{P}\,\hat{G}^\lambda_{\mathrm{shell}}]
  =
  \int_0^\infty \frac{d k^0}{2\pi} \mathrm{Tr} \big[\mathcal{P} ( \hat{g}^>\hat{G}^<_{\mathrm{shell}} - \hat{g}^<\hat{G}^>_{\mathrm{shell}} ) \big].
  \label{eq:projected_collision_term}
\end{align}
Combining this with the Liouville term, the Boltzmann equation is obtained from Eq.~\eqref{eq:Liouville_vs_collision} as
\begin{align}
  \int_0^\infty \frac{d k^0}{2\pi}
  H[\mathcal{P}\,\hat{G}^\lambda_{\mathrm{shell}}]
  =
  \int_0^\infty \frac{d k^0}{2\pi}
  \mathrm{Tr}
  \big[\mathcal{P} ( \hat{g}^>\hat{G}^<_{\mathrm{shell}}
  -
  \hat{g}^<\hat{G}^>_{\mathrm{shell}} )
  \big].
  \label{eq:projected_boltzmann_master}
\end{align}
The explicit form of the Liouville term on the left-hand side depends on the structure of the on-shell solution and is therefore evaluated separately for bosons and fermions in the following. 

\subsubsection{Bosons}

For the bosonic case, substituting Eq.~\eqref{eq:boson_Wightman_shell} into the integrand of the left-hand side in Eq.~\eqref{eq:projected_boltzmann_master}, we obtain
\begin{align}
  H[P_i \, i \hat{\Delta}^\lambda_{\mathrm{shell}}]&= 2 \pi \mathrm{sign}(k^0) \{\Omega_{i}^2 , n^\lambda_{i} \}_{z,k_z} \delta(\Omega_{i}^2).
\end{align}
By integrating the equation for the lesser function ($\lambda = <$) over $k^0 > 0$, we obtain the Liouville term as 
\begin{align}
  \bm{L}_i[f_{i}] \equiv \int_0^\infty \frac{d k^0}{2 \pi} H[P_i \, i \hat{\Delta}^<_{\mathrm{shell}}] &= \frac{1}{|\partial_{k^0} \Omega_i^2|} \big\{ \Omega_i^2 , n_i^< \big\}_{z,k_z} \bigg|_{k^0 = \omega_i} \notag \\
   &= \partial_{k_z} \omega_i \partial_{z} f_i - \partial_z \omega_i \partial_{k_z} f_i \notag \\
  &= \dot{z}_i \partial_z f_i + \dot{p}_{z,i} \partial_{k_z} f_i,
  \label{eq:boson_liouville}
\end{align}
where $k^0 = + \omega_i$ is a positive-energy root of $\Omega_{i}^2 (k^0) = 0$.
Here, we focus on a single isolated positive-energy quasiparticle branch.
We have defined $f_i(\bm{k}_\parallel,k_z,z)=n_i^<(k^0=\omega_i,\bm{k}_\parallel,k_z,z)$ and used the canonical relations
$\dot z_i=\partial_{k_z}\omega_i$ and $\dot p_{z,i}=-\partial_z \omega_i$ in the last line.
Here, $\dot z_i$ is the group velocity of the bosonic $i$-th flavor quasiparticle along the $z$ direction, while $\dot p_{z,i}$ is the semiclassical force acting on it.

From Eq.~\eqref{eq:projected_boltzmann_master}, the collision term for the bosonic case is given by 
\begin{align}
  \bm{C}_i[f_i,\cdots] \equiv \int_0^\infty \frac{d k^0}{2\pi}
  \mathrm{Tr}
  \big[P_i ( \hat{\Pi}^> \hat{\Delta}^<_{\mathrm{shell}}
  -
  \hat{\Pi}^< \hat{\Delta}^>_{\mathrm{shell}} )
  \big],
  \label{eq:boson_coll}
\end{align}
where the trace is taken in flavor space.
This expression is general and still written in terms of the Wightman self-energy.
Evaluating the Wightman self-energy with the on-shell projection for the internal particles, one can obtain the explicit form of the collision term.
However, we do not show its explicit form here, since it depends on the model and interactions (e.g. see ref.~\cite{Prokopec:2004ic}).

Combining Eqs.~\eqref{eq:boson_liouville} and \eqref{eq:boson_coll}, we obtain the Boltzmann equation for the bosonic field,
\begin{align}
  \bm{L}_i[f_i] = \bm{C}_i[f_i,\cdots].
\end{align}
In the conventional analysis without self-energy corrections, the first-order correction to the bosonic quasiparticle shell does not generate a CP-violating semiclassical force~\cite{Kainulainen:2001cn,Prokopec:2003pj}.
Here we have the shell function $\Omega_i^2$ corrected by the Hermitian part of the self-energy.
Even in this formulation, no CP-odd correction to the quasiparticle shell appears for bosons at this order, unlike in the fermionic case discussed below.
Thus, the dispersive self-energy correction modifies the quasiparticle dispersion relation without generating an additional CP-odd force term.
In the limit $\Pi_{\mathcal H}\to0$, our result reduces to the conventional semiclassical Boltzmann equation of refs.~\cite{Kainulainen:2001cn,Prokopec:2003pj}.

\subsubsection{Fermions}

For the fermionic case, the derivation proceeds in parallel with the bosonic case.
By substituting Eq.~\eqref{eq:fermion_shell_n1} into the integrand of the left-hand side in Eq.~\eqref{eq:projected_boltzmann_master}, after a straightforward calculation, we obtain
\begin{align}
  H[P_s P_i \, i \hat{S}^\lambda_{\mathrm{shell}}]&= 2 \pi \mathrm{sign}(\tilde{k}^0) \{\Omega_{s,i,\mathrm{eff}}^2 , n^\lambda_{s,i} \}_{z,k_z} \delta(\Omega_{s,i,\mathrm{eff}}^2),
  \label{eq:fermion_hermitian}
\end{align}
where 
\begin{align}
  \delta \Omega_{s,i}^2 &= -\frac{1}{\tilde{\mathcal{K}}^0} \bigg[m_{Ii}^\prime(\mathcal{K}_z \partial_{k_z} \mathcal{M}_{Ri} -  \mathcal{M}_{Ri} \partial_{k_z} \mathcal{K}_z ) - m_{Ri}^\prime(\mathcal{K}_z \partial_{k_z} \mathcal{M}_{Ii} -  \mathcal{M}_{Ii} \partial_{k_z} \mathcal{K}_z ) \bigg], \notag \\
  \Omega_{s,i,\mathrm{eff}}^2 &= \Omega_{s,i}^2 + \delta \Omega_{s,i}^2.
  \label{eq:corrected_shell}
\end{align}
Although the spin index $s$ does not appear explicitly on the right-hand side of $\delta\Omega_{s,i}^2$, the spin dependence is carried by $\mathcal{M}_{Ri}$, $\mathcal{M}_{Ii}$, $\mathcal{K}_z$, and $\tilde{\mathcal{K}}^0$, as defined in Eq.~\eqref{eq:def_coef_fer}.
By integrating Eq.~\eqref{eq:fermion_hermitian} for the lesser function over $\tilde{k}^0 > 0$, the Liouville term is obtained by
\begin{align}
  \bm{L}_{s,i}[\tilde{f}_{s, i}] \equiv \int_0^\infty \frac{d \tilde{k}^0}{2 \pi} H[P_s P_i \, i \hat{S}^<_{\mathrm{shell}}] &= 
  \frac{1}{|\partial_{\tilde{k}^0} \Omega_{s,i,\mathrm{eff}}^2|} \big\{ \Omega_{s,i,\mathrm{eff}}^2 , n_{s,i}^< \big\}_{z,k_z} \bigg|_{\tilde{k}^0 = \tilde{\omega}_{s,i}} \notag \\
  &= \partial_{k_z} \tilde{\omega}_{s,i} \partial_{z} \tilde{f}_{s,i} - \partial_z \tilde{\omega}_{s,i} \partial_{k_z} \tilde{f}_{s,i} \notag \\
  &= \dot{z}_{s,i} \partial_z \tilde{f}_{s,i} + \dot{p}_{z,s,i} \partial_{k_z} \tilde{f}_{s,i},
  \label{eq:fermion_liouville_boost}
\end{align}
where $\tilde{k}^0 = + \tilde{\omega}_{s,i}$ is a positive-energy root of $\Omega_{s,i,\mathrm{eff}}^2 = 0$.
As in the bosonic case, for each spin projection $s$ and flavor index $i$, we focus on an isolated positive-energy quasiparticle branch.
To distinguish the distribution function in the boosted frame, we have defined $\tilde{f}_{s, i} (k_z, z) = n^<_{s,i} (\tilde{k}^0 = \tilde{\omega}_{s,i}, k_z, z)$.
We have used $\dot z_{s,i}=\partial_{k_z}\tilde\omega_{s,i}$ and
$\dot p_{z,s,i}=-\partial_z\tilde\omega_{s,i}$.
Again, $\dot z_{s,i}$ is the group velocity of the fermionic $i$-th flavor quasiparticle with the spin $s$ along the $z$ direction, while $\dot p_{z,s,i}$ is the semiclassical force acting on it.

From Eq.~\eqref{eq:projected_boltzmann_master}, the collision term is given by
\begin{align}
  \bm C_{s,i} [\tilde{f}_{s,i},\cdots] \equiv 
  \int_0^\infty \frac{d\tilde{k}^0}{2\pi}
  \mathrm{Tr}
  \big[
  P_s P_i
  \big(
  \hat{\Sigma}^> \hat{S}^<_{\mathrm{shell}}
  -
  \hat{\Sigma}^< \hat{S}^>_{\mathrm{shell}}
  \big)
  \big],
  \label{eq:fermionic_collision_general}
\end{align}
where the trace is taken over spinor and flavor spaces.
Its explicit form is model-dependent and is obtained by evaluating $\hat\Sigma^\lambda$ for the relevant scattering and decay processes, with the internal particles put on-shell.
For explicit examples, see refs.~\cite{Prokopec:2004ic,Kainulainen:2021oqs}.

Equating Eqs.~\eqref{eq:fermion_liouville_boost} and \eqref{eq:fermionic_collision_general}, the fermionic Boltzmann equation in the boosted frame is written as
\begin{align}
  \bm L_{s,i}[\tilde{f}_{s,i}]
  =
  \bm C_{s,i}[\tilde{f}_{s,i},\cdots].
  \label{eq:fermionic_Boltzmann_boosted}
\end{align}
So far, the fermionic Boltzmann equation has been derived in the boosted frame.
However, the fluid ansatz for the distribution function given in Eq.~\eqref{eq:distribution_wall_perturbation}, which is conventionally used in EWBG calculations~\cite{Huet:1995sh,Cline:2000nw,Kainulainen:2002th,Prokopec:2003pj,Prokopec:2004ic,Fromme:2006wx,Cline:2020jre,Kainulainen:2021oqs,Kainulainen:2024qpm}, is formulated in the wall frame.
We therefore rewrite Eq.~\eqref{eq:fermionic_Boltzmann_boosted} in the wall frame below.

We emphasize again that the spinor decomposition of the fermionic functions has been performed in the boosted frame where $\tilde{\bm{k} }_\parallel=0$.
Therefore, $a_{0s}^{r,a,\lambda}$ and $a_{ms}^{r,a,\lambda}$ appearing in $\Omega^2_{s,i,\mathrm{eff}}$ are boosted-frame coefficients which are evaluated at $\tilde{k}^\mu = \Lambda_\parallel^\mu {}_\nu k^\nu$ and $\tilde{u}^\mu = \Lambda_\parallel^\mu {}_\nu u^\nu$.
We define the effective shell function in the wall frame by
\begin{align}
  \Omega^2_{s,i,\mathrm{eff},w}
  (k^0,\bm{k}_\parallel,k_z,z;u)
  \equiv
  \Omega^2_{s,i,\mathrm{eff}}
  (\tilde k^0,k_z,z;\tilde u).
  \label{eq:Omega_eff_wall_def}
\end{align}
The positive-energy solution in the wall frame is defined by
\begin{align}
  \Omega^2_{s,i,\mathrm{eff},w}
  (k^0=\omega_{s,i},\bm{k}_\parallel,k_z,z;u)
  =
  0 .
\end{align}

Rewriting Eq.~\eqref{eq:fermion_hermitian} in terms of $\Omega^2_{s,i,\mathrm{eff},w}$ and integrating over $k^0 > 0$, we obtain the Liouville term in the wall frame as 
\begin{align}
  \bm L^{w}_{s,i}[f_{s,i}] = \int_0^\infty dk^0 \{\Omega_{s,i,\mathrm{eff},w}^2 , n^<_{s,i} \}_{z,k_z} \delta(\Omega_{s,i,\mathrm{eff},w}^2) &= \frac{1}{|\partial_{k^0}
  \Omega^2_{s,i,\mathrm{eff},w}|}
  \big\{
  \Omega^2_{s,i,\mathrm{eff},w},
  n^<_{s,i}
  \big\}_{z,k_z}
  \bigg|_{k^0=\omega_{s,i}}
  \notag \\
  &=\partial_{k_z}\omega_{s,i}\,\partial_z f_{s,i}
  -\partial_z\omega_{s,i}\,\partial_{k_z}f_{s,i}
  \notag \\
  &=
  \dot z_{s,i}\,\partial_z f_{s,i}
  +
  \dot p_{z,s,i}\,\partial_{k_z}f_{s,i}.
  \label{eq:fermion_liouville_wall}
\end{align}
We have defined
$f_{s,i}(\bm{k}_\parallel,k_z,z) = n^<_{s,i}(k^0=\omega_{s,i},\bm{k}_\parallel,k_z,z)$.
From Eq.~\eqref{eq:projected_boltzmann_master}, the collision term in the wall frame is obtained as
\begin{align}
  \bm{C}_{s,i}^w [f_{s,i},\cdots] = \int_0^\infty \frac{d k^0}{2\pi} \mathrm{Tr} \big[ P_s P_i \big( \hat{\Sigma}^> \hat{S}^<_{\mathrm{shell}} - \hat{\Sigma}^< \hat{S}^>_{\mathrm{shell}} \big) \big].
  \label{eq:fermionic_collision_general_wall}
\end{align}
Thus, we arrive at the fermionic Boltzmann equation in the wall frame,
\begin{align}
  \bm L_{s,i}^w [f_{s,i}]
  =
  \bm C_{s,i}^w [f_{s,i},\cdots].
  \label{eq:fermionic_Boltzmann_wall}
\end{align}

We conclude this section by discussing the Boltzmann equation for the antiparticle in the wall frame.
The Boltzmann equation for the antiparticle is obtained by evaluating Eq.~\eqref{eq:fermionic_Boltzmann_wall} in the CP-conjugated wall and plasma backgrounds.
In particular, if the complex phase of the mass parameter is the only source of CP violation, $\overline{a}_{2s,\mathcal H}$ obeys the same CP transformation as $m_{Ii}$.
Therefore, the imaginary part of the effective mass, $\mathcal{M}_{Ii}=m_{Ii}+\overline{a}_{2s,\mathcal H}$, is CP odd, and hence $\delta\Omega_{s,i}^2$ changes sign under CP conjugation\footnote{
    If the plasma itself is CP violating, a more general model-dependent analysis is required.
    In the electroweak plasma, gauge interactions preserve chirality.
    Therefore, the chiral-symmetry-breaking components of the self-energy, $a_1$ and $a_2$, can arise only through chirality-changing interactions or mass insertions and are expected to be suppressed.
}.
The effective shell function for the antiparticle is therefore obtained as
\begin{align}
  \Omega^{2}_{s,i,\mathrm{eff}} \Big|_{\mathrm{CP}} = \Omega^2_{s,i} - \delta\Omega^2_{s,i}.
\end{align}
Replacing $\Omega^{2}_{s,i,\mathrm{eff},w} \to \Omega^{2}_{s,i,\mathrm{eff},w} \big|_{\mathrm{CP}}$ in Eq.~\eqref{eq:fermion_liouville_wall}, we obtain the Liouville term for the antiparticle in the wall frame as
\begin{align}
  \bm L^{w,\mathrm{CP}}_{s,i}[f_{s,i}^{\mathrm{CP}}]
  &=
  \partial_{k_z}\omega_{s,i}^{\mathrm{CP}}\,\partial_z f_{s,i}^{\mathrm{CP}}- \partial_z\omega_{s,i}^{\mathrm{CP}}\,\partial_{k_z}f_{s,i}^{\mathrm{CP}},
\end{align}
where $k^0=\omega_{s,i}^{\mathrm{CP}}$ satisfies $\Omega^{2}_{s,i,\mathrm{eff},w}\big|_{\mathrm{CP}} = 0$, and $f_{s,i}^{\mathrm{CP}} = n^<_{s,i}(k^0=\omega_{s,i}^{\mathrm{CP}},\bm{k}_\parallel,k_z,z)$.
The difference between $\omega_{s,i}^{\mathrm{CP}}$ and $\omega_{s,i}$ gives rise to CP-odd differences in the group velocity and semiclassical force terms.

Compared with ref.~\cite{Kainulainen:2021oqs}, our expression retains $\overline{a}_{1s, \mathcal{H}}$ and $\overline{a}_{2s, \mathcal{H}}$ in the effective mass components $\mathcal{M}_{R i}$ and $\mathcal{M}_{I i}$, respectively, when constructing the effective shell function $\Omega^2_{s,i,\mathrm{eff}}$.
If these contributions are neglected, our result is consistent with the thermally corrected semiclassical Boltzmann equation derived in ref.~\cite{Kainulainen:2021oqs}.
If all self-energy corrections are further neglected, the Boltzmann equation reproduces the conventional semiclassical result of refs.~\cite{Kainulainen:2001cn,Prokopec:2003pj,Kainulainen:2002th}.

\section{Conclusions \label{sec:conclusions}}

In this paper, we systematically derived the Boltzmann equation relevant for electroweak baryogenesis from the Kadanoff--Baym equations while retaining the self-energy corrections. 
We solved the Kadanoff--Baym equations algebraically and order by order in the derivative expansion. 
We then clarified the conditions to obtain the Boltzmann equation for electroweak baryogenesis calculations.

Within this framework, we identified the on-shell part of the spectral and Wightman functions for both bosonic and fermionic systems. 
The Hermitian part of the self-energy modifies the quasiparticle shell, while the anti-Hermitian part determines the width of the spectral peak around the quasiparticle shell.
In the narrow-width limit, this peak reduces to the on-shell delta-function contribution. 
We also introduced a decomposition of the Wightman self-energy into a part proportional to the product of the statistical function and the absorptive part of the self-energy, which generates the on-shell part of the Wightman function, and a remaining interaction-dependent part contributing to the collision term.
After projecting the Hermitian part of the Kadanoff--Baym equation onto the quasiparticle shell, we obtained the Boltzmann equation for the corresponding flavor-diagonal distribution function.

For bosons, the self-energy modifies the quasiparticle shell at zeroth order in the derivative expansion.
Even so, under the assumptions adopted in this work, no additional flavor-diagonal correction to the shell arises at first order.
Consequently, no CP-odd correction to the group velocity or the semiclassical force appears at this order.
For fermions, by contrast, the first-order correction to the spectral function contains a derivative of the delta function and can be interpreted as a shift of the quasiparticle shell. 
This shift modifies both the velocity and force terms in the Liouville term. 
In the presence of a spacetime-dependent complex mass, it contains a CP-odd contribution and therefore leads to different propagation for particles and antiparticles. 
Our result reproduces the conventional semiclassical Boltzmann equation if the self-energy corrections are neglected.

Several extensions remain for future work. 
In this paper, we assumed that the dominant plasma corrections are flavor universal and restricted the Boltzmann equations to the flavor-diagonal elements. 
Quantum coherence between different flavor modes can be incorporated by retaining the off-diagonal components of the propagators and self-energy in the local mass basis, which is particularly important for nearly degenerate masses~\cite{Konstandin:2004gy,Konstandin:2005cd,Herranen:2008hi,Cirigliano:2009yt,Cirigliano:2011di}. 
It would also be useful to include the spacetime dependence and more general spin, chiral, and CP-violating structures of the self-energy, and to examine cases in which the simple quasiparticle approximation used here would be insufficient. 
Finally, applications to specific models of electroweak baryogenesis require evaluating the dispersive and absorptive self-energies, constructing the corresponding collision terms, specifying the bubble-wall background, and solving the resulting transport equations.
The formulation developed in this work provides a systematic framework for these extensions.

\section*{Acknowledgements}
This work was supported by JSPS KAKENHI Grant Numbers 22K21347 (M.E. and Y.M.) and 26KJ1238 (T.T.).

\appendix

\section{Discussion of the VEV-insertion approximation}
\label{sec:VIA}

In this appendix, we discuss the VEV-insertion approximation (VIA)~\cite{Riotto:1995hh,Riotto:1997vy,Riotto:1998zb}, which provides the transport equations for EWBG in a different way from the semiclassical approach~\cite{Kainulainen:2001cn,Kainulainen:2002th,Prokopec:2003pj}.

It has been recognized that the CP-violating source term derived in the VIA differs in form from that in the semiclassical approach, leading to significantly different predictions for the baryon asymmetry~\cite{Cline:2020jre,Cline:2021dkf,Basler:2021kgq}.
These discrepancies have motivated a closer examination of the VIA, and several issues have been raised~\cite{Postma:2019scv,Postma:2021zux,Kainulainen:2021oqs,Postma:2022dbr,vandeVis:2025efm}.
In particular, ref.~\cite{Kainulainen:2021oqs} identified problems associated with treating the spacetime-dependent mass as an insertion in a nonlocal self-energy, while ref.~\cite{Postma:2022dbr} argued that the conventional VIA source vanishes due to a cancellation between the mass-commutator and collision contributions.

In the following, we first briefly review the derivation of the VIA source term~\cite{Riotto:1995hh,Riotto:1997vy,Riotto:1998zb} and discuss the issue raised in ref.~\cite{Kainulainen:2021oqs}.
We then revisit the cancellation of the VIA source claimed in ref.~\cite{Postma:2022dbr} and clarify its meaning from the perspective of the analytic solution derived in section~\ref{sec:solution}.

\subsection{The VIA source term and pinch singularity}

The conventional VIA developed in refs.~\cite{Riotto:1995hh,Riotto:1997vy,Riotto:1998zb} takes as its starting point the divergence equation for a current associated with the transported charge.
Since this equation directly provides the continuity equation for the corresponding charge, a CP-odd contribution to the current divergence is identified as a source term in the transport equation.

To illustrate this construction, we consider in this subsection a single Dirac fermion $\psi$ with a spacetime-dependent complex mass $m(u)=|m(u)|e^{i\theta(u)}$.
The current for the left-handed component is defined by\footnote{
    In the fermionic derivations of refs.~\cite{Riotto:1997vy,Riotto:1998zb}, the current-divergence equation is formulated in terms of the greater Wightman function $S^>$.
    Here, following ref.~\cite{Kainulainen:2021oqs}, we instead use the lesser Wightman function $S^<$ with the convention adopted in this work.
    The choice of the Wightman component does not affect the discussion below.
}
\begin{align}
  j^\mu_{L}(u) \equiv \langle\bar{\psi}(u)\gamma^\mu P_{L} \psi(u)\rangle
  =
  -\mathrm{Tr}\!\left[\gamma^\mu P_{L}\,i S^<(u,u)\right],
  \label{eq:VIA_currentdif_fermion}
\end{align}
where $\mu$ is the Lorentz index.
The right-handed current is defined analogously by replacing $L\to R$.
The continuity equation for $j^\mu_L$ is schematically written as
\begin{align}
  \partial_\mu j_L^\mu
  =
  \partial_t j^0_L+\bm{\nabla}\cdot\bm{j}_L
  =
  S_{L,\mathrm{VIA}}^{\CPV}+\mathcal{C}_L.
  \label{eq:VIA_current_transport}
\end{align}
Here, $S_{L,\mathrm{VIA}}^{\CPV}$ denotes the CP-violating VIA source for the left-handed charge, while $\mathcal{C}_L$ denotes the remaining contributions from processes in the plasma that relax the corresponding charge.
The spatial current is related to the number density through Fick's law, $\bm{j}_L=-D_L\bm{\nabla}j_L^0$, where $D_L$ is the diffusion constant for the left-handed fermion.
The continuity equation in Eq.~\eqref{eq:VIA_current_transport} is then converted into a transport equation for the number density $j_L^0$.
Combining Eq.~\eqref{eq:SDequation2} with its conjugate equation, we obtain the explicit form of Eq.~\eqref{eq:VIA_current_transport}:
\begin{align}
    \partial_\mu j_L^\mu(u) ={}& -\lim_{v \to u} \mathrm{Tr}\!
    \left[ \left( m(u) P_R - m^\ast(v) P_L \right) S^<(u,v) \right]
\notag\\
&-2 \, \int d^3\bm{w} \int_{- \infty}^{u^0}dw^0\,
\mathrm{Re} \bigg[ \mathrm{Tr}\!\left[ P_L \left( \Sigma^>(u,w) S^<(w,u) - \Sigma^<(u,w) S^>(w,u) \right) \right] \bigg].
\label{eq:exact_currentdivergence}
\end{align}
The first term on the right-hand side contains the contribution from the local spacetime-dependent mass operator, whereas the second term is a nonlocal memory integral involving the Wightman self-energies.

In the conventional VIA calculations in refs.~\cite{Riotto:1995hh,Riotto:1997vy,Riotto:1998zb}, the first term in Eq.~\eqref{eq:exact_currentdivergence} is not retained in evaluating the VIA source.
Instead, the effect of the spacetime-dependent mass is incorporated into the second term through a nonlocal self-energy contribution.
At the lowest nontrivial order retained in the conventional VIA, involving two mass (vacuum) insertions, we have
\begin{align}
    &\mathrm{Tr}
    \left[ P_L \left( \Sigma^>(u,w) S^<(w,u) - \Sigma^<(u,w) S^>(w,u) \right)
    \right]
    \notag \\
    \simeq{}&
    \mathrm{Tr} \left[ \Sigma_{L,\mathrm{VIA}}^>(u,w) S^<_{0,L}(w,u) - \Sigma^<_{L,\mathrm{VIA}}(u,w) S^>_{0,L}(w,u) \right],
    \label{eq:VIA_LOtrace}
\end{align}
where the left-handed Wightman self-energy is given by
\begin{align}
  i\Sigma_{L,\mathrm{VIA}}^\lambda(u,w) =
  [-im(u)]\, iS^\lambda_{0,R}(u,w)\,
  [-im^\ast(w)].
  \label{eq:VIA_selfenergy}
\end{align}
The propagators $S^\lambda_{0,L/R}$ are the massless free Wightman propagators for $\psi_{L/R}$ evaluated with equilibrium thermal distribution functions.

Substituting Eqs.~\eqref{eq:VIA_LOtrace} and \eqref{eq:VIA_selfenergy} into Eq.~\eqref{eq:exact_currentdivergence} and retaining the CP-odd part, we obtain the CP-violating VIA source
\begin{align}
  S_{L,\mathrm{VIA}}^{\CPV}(u)
  ={}&
  2|m(u)|^2\partial_{\mu}\theta(u)
  \int d^4w\,
  \Theta(u^0-w^0)\,
  (u-w)^\mu \times I_{\mathrm{VIA}}(u,w) + \mathcal{O}(\partial_u^2),
  \label{eq:VIA_CP_source}
\end{align}
where 
\begin{align}
    I_{\mathrm{VIA}}(u,w) = \mathrm{Im}\bigg[ \mathrm{Tr} \left[ iS_{0,L}^>(u,w)iS_{0,R}^<(w,u) - iS_{0,L}^<(u,w)iS_{0,R}^>(w,u) \right] \bigg].
    \label{eq:VIA_integral}
\end{align}
We have expanded the mass as
$m(w)=m(u)+(w-u)^\mu\partial_{\mu}m(u)+\mathcal{O}(\partial_u^2)$.
The corresponding contribution to the right-handed current has the opposite sign,
\begin{align}
  S_{R,\mathrm{VIA}}^{\CPV} = -S_{L,\mathrm{VIA}}^{\CPV}.
  \label{eq:VIA_chiral_source_relation}
\end{align}
Thus, the CP-violating VIA source cancels in the vector-current divergence and contributes only to the axial-current divergence.
The resulting chiral asymmetry can bias electroweak sphaleron transitions in the symmetric phase, thereby contributing to the generation of the baryon asymmetry.

Several problems with this procedure were pointed out in ref.~\cite{Kainulainen:2021oqs}.
In particular, the VIA source contains products of propagators with overlapping quasiparticle poles.
After transforming the propagator products in Eq.~\eqref{eq:VIA_integral} to momentum space, the relevant momentum integral schematically takes the form
\begin{align}
  I_{\mathrm{VIA}} \sim \int\frac{d^4k}{(2\pi)^4} k^2 (\mathrm{sign}(k^0) )^2 \delta(k^2)^2.
  \label{eq:VIA_pinch}
\end{align}
The overlap of the quasiparticle poles produces the squared delta function, which is not well defined as a distribution and constitutes the pinch singularity.
In the conventional VIA calculations~\cite{Riotto:1995hh,Riotto:1997vy,Riotto:1998zb}, the zero-width propagators are replaced by dressed propagators containing finite thermal widths, thereby regularizing the explicit pinch singularity.
However, ref.~\cite{Kainulainen:2021oqs} showed that introducing finite thermal widths does not uniquely define the VIA source.
Since the zero-width expression is itself ill-defined, its extension to finite width is not unique, and different extensions can give different finite results.
Thus, finite thermal widths regularize the pinch singularity without resolving the underlying ambiguity of the VIA source.

More fundamentally, ref.~\cite{Kainulainen:2021oqs} identified the central problem of the conventional VIA; the local spacetime-dependent mass is used in the construction of the nonlocal self-energy (cf. Eq.~\eqref{eq:VIA_selfenergy}).
Thus, the corresponding self-energy diagram is not one-particle irreducible.
By contrast, in the Kadanoff--Baym equations used in the semiclassical treatment, as in refs.~\cite{Kainulainen:2001cn,Kainulainen:2002th,Prokopec:2003pj}, the spacetime-dependent mass is included in the mass term appearing in the inverse propagator, so that its effects are consistently resummed into the propagators.
The wall-gradient effects are then incorporated through the derivative expansion.

\subsection{Revisiting the VIA cancellation}

Another claim concerning the VIA was made in ref.~\cite{Postma:2022dbr}.
Their analysis considered both a two-flavor bosonic system and a single-flavor fermionic system.
They first performed the Wigner transformation and then related the momentum-integrated kinetic equation derived from Eq.~\eqref{eq:KBeq_1} to a current-divergence equation.
The corresponding equation can also be derived directly without performing the Wigner transformation, as in refs.~\cite{Riotto:1995hh,Riotto:1997vy,Riotto:1998zb} (cf. Eq.~\eqref{eq:exact_currentdivergence}).
To evaluate the VIA source term, they solved the constraint equations at zeroth order in the derivative expansion, both perturbatively in the mass insertions and with the mass insertions fully resummed.
They then substituted these solutions into the current-divergence equation.
They found that the source term vanishes due to an exact cancellation between the mass-commutator and collision contributions.
They therefore concluded that the conventional VIA source vanishes at leading order in the derivative expansion.
To understand their analysis from the perspective of our formulation, we revisit their calculation below, following the order of their discussion.

Defining
\begin{align}
  \mathcal{I}^\lambda \equiv - [ \mathcal{D}_C, G^\lambda] + e^{-i\diamond} \Big( [\mathcal{M}+g_{\mathcal H}, G^\lambda ] + [g^\lambda, G_{\mathcal H}] \Big) + \frac{1}{2} e^{-i\diamond} \Big( \{g^>,G^<\} - \{g^<,G^>\} \Big),
  \label{eq:VIA_I_definition}
\end{align}
Eq.~\eqref{eq:KBeq_1} can be written as
\begin{align}
  \big\{ \mathcal{D}_K, G^\lambda \big\} &= \mathcal{I}^\lambda.
  \label{eq:VIA_kinetic}
\end{align}
Expanding $e^{-i \diamond}$ and $\mathcal{D}_C$ in derivatives and retaining the zeroth-order terms, we obtain
\begin{align}
  \left. \mathcal{I}^{\lambda} \right|_{\diamond^0} &= - [ \mathcal{D}^{[0]}, G^{\lambda}] + [\mathcal{M}+g_{\mathcal H}, G^{\lambda} ] + [g^\lambda, G_{\mathcal H} ] + \frac{1}{2} \Big( \{g^>,G^{<} \} - \{g^<,G^{>}\} \Big).
  \label{eq:VIA_I_diamond0}
\end{align}
Equation~\eqref{eq:VIA_I_diamond0} keeps explicit the mass-commutator and collision contributions discussed in ref.~\cite{Postma:2022dbr}, while the algebraically equivalent form below makes the subsequent VIA cancellation more transparent:
\begin{align}
  \left. \mathcal{I}^{\lambda} \right|_{\diamond^0} = -\alpha_-G^{\lambda} +g^\lambda G^{a} +G^{\lambda}\alpha_+ -G^{r}g^\lambda.
  \label{eq:VIA_I_diamond0_alt}
\end{align}

For the two-flavor bosonic system ($i = 1,2$)\footnote{
    In ref.~\cite{Postma:2022dbr}, the two flavors are denoted by $i = L,R$.
}, we project Eq.~\eqref{eq:VIA_kinetic} onto the first flavor component and integrate over momentum.
Retaining the zeroth-order contribution to the right-hand side in Eq.~\eqref{eq:VIA_I_diamond0}, we obtain
\begin{align}
    \partial_{x,\mu} J_1^{\mu,\lambda}(x)  = S_{\mathrm{B}, 11}^\lambda.
    \label{eq:VIA_bosonic_currentdiv}
\end{align}
The left-hand side of this equation is given by
\begin{align}
  \partial_{x, \mu} J_1^{\mu, \lambda}(x) = \int \frac{d^4k}{(2\pi)^4} \mathrm{Tr} \Big[ P_1 \{ i k \cdot \partial_x , \Delta^\lambda \} \Big],
  \label{eq:VIA_boson_currentlhs}
\end{align}
while the right-hand side is given by
\begin{align}
  S_{\mathrm{B}, 11}^{\lambda} = \int \frac{d^4k}{(2\pi)^4} \mathrm{Tr} \Big[ P_1 \left. \mathcal{I}^\lambda \right|_{\diamond^0} \Big],
  \label{eq:VIA_boson_source}
\end{align}
where $P_1=\mathrm{diag}(1,0)$ is the projection matrix onto the first component in flavor space, and $G=\Delta$ and $g=\Pi$ are used in $\mathcal{I}^\lambda$.
Taking the trace with the projection matrix $P_1$ selects the diagonal $(1,1)$ component of the two-by-two matrix in flavor space.
For $\lambda = <$, the quantity $J_1^{\mu,<}(x)$ coincides with the bosonic current,\footnote{
    In Eq.~(3.5) of ref.~\cite{Postma:2022dbr}, $\Delta^<$ and $\Delta^>$ are summed in relating the kinetic equation to the current divergence.
    By contrast, the bosonic current-divergence equation in ref.~\cite{Riotto:1998zb}, cited there, is derived from $\Delta^<$ alone.
    Moreover, directly combining Eqs.~(3.4)-(3.6) of ref.~\cite{Postma:2022dbr} appears to lead to an overall normalization differing by a factor of four from their Eq.~(3.5).
    This normalization issue is irrelevant to the VIA cancellation discussed below, since the cancellation holds separately for each Wightman function.
}
\begin{align}
    J^{\mu,<}_{1} (x) = \langle \phi_1^\dagger(v) (i \overrightarrow{\partial_u} - i \overleftarrow{\partial_v} )^\mu \phi_1 (u) \rangle \Big|_{u=v=x}.
\end{align}
Therefore, Eq.~\eqref{eq:VIA_bosonic_currentdiv} with $\lambda = <$ gives the transport equation for $\phi_1$.
After summing Eq.~\eqref{eq:VIA_boson_source} over $\lambda=<,>$, the resulting expression is equivalent to the negative of Eq.~(3.6) of ref.~\cite{Postma:2022dbr}.

For the single-flavor fermionic system, we decompose the matrix products in Eq.~\eqref{eq:VIA_kinetic} into chiral blocks and retain their $LL$ components before taking the trace over spin space.
We define the $LL$ component of a product in chiral space by
\begin{align}
  (A B)_{LL} \equiv \sum_{I=L,R} A_{LI} B_{IL}.
  \label{eq:VIA_LL_product}
\end{align}
For the propagator, the $LL$ component is given by $S_{LL}=P_L S P_R$, whereas for the mass and self-energy, the corresponding components are given by $\mathcal{M}_{LL}=P_R \mathcal{M} P_L$ and $\Sigma_{LL}=P_R \Sigma P_L$, respectively.
Projecting Eq.~\eqref{eq:VIA_kinetic} onto the $LL$ component, integrating over momentum, and retaining the zeroth-order contribution to the right-hand side in Eq.~\eqref{eq:VIA_I_diamond0}, we obtain
\begin{align}
    \partial_{x,\mu} j_L^{\mu,\lambda}(x) = S_{\mathrm{F},LL}^{\lambda}.
    \label{eq:VIA_fermionic_currentdiv}
\end{align}
The left-hand side of this equation is given by
\begin{align}
  \partial_{x, \mu} j_L^{\mu, \lambda}(x) = - \int \frac{d^4k}{(2\pi)^4} \mathrm{Tr} \Big[ \Big\{ \frac{i}{2} \cancel{\partial}_x,  P_L S^\lambda \Big\} \Big],
  \label{eq:VIA_fermion_current}
\end{align}
while the right-hand side is given by
\begin{align}
  S_{\mathrm{F}, LL}^{\lambda} = - \int \frac{d^4k}{(2\pi)^4} \mathrm{Tr} \Big[ \Big( \left. \mathcal{I}^\lambda \right|_{\diamond^0} \Big)_{LL} \Big].
  \label{eq:VIA_fermion_source}
\end{align}
Here, $G=S$ and $g=\Sigma$ are used in $\mathcal{I}^\lambda$.
The trace in Eq.~\eqref{eq:VIA_fermion_current} is taken over the four-dimensional Dirac spinor space, whereas that in Eq.~\eqref{eq:VIA_fermion_source} is taken over the two-dimensional spin space after extracting the $LL$ component in chiral space.
For the lesser Wightman function, $\lambda=<$, the quantity $j_L^{\mu,<}(x)$ coincides with the current associated with $\psi_L$\footnote{
    In ref.~\cite{Postma:2022dbr}, the Wightman component $\lambda$ used in defining the current is left unspecified.
    Here, we use the lesser Wightman function to define the current.
    This difference is irrelevant to the VIA cancellation discussed below.
}
\begin{align}
    j^{\mu,<}_{L} (x) = \langle \overline{\psi_L}(v) \gamma^\mu \psi_L (u) \rangle \Big|_{u=v=x}.
\end{align}
We note that this expression is equivalent to Eq.~\eqref{eq:VIA_currentdif_fermion}.
After summing Eq.~\eqref{eq:VIA_fermion_source} over $\lambda=<,>$, the resulting expression is equivalent to the negative of Eq.~(4.6) of ref.~\cite{Postma:2022dbr}.

The zeroth-order solution given in Eq.~\eqref{eq:sol_n0} provides a general expression that includes the fully resummed solutions obtained in ref.~\cite{Postma:2022dbr} as special cases.
In particular, upon specializing to the two-flavor bosonic and single-flavor fermionic systems considered there, the solution for the Wightman functions, $G^{\lambda,[0]}=\alpha_-^{-1}g^\lambda\alpha_+^{-1}$, reproduces the solutions given in Eqs.~(3.30) and (4.29) of ref.~\cite{Postma:2022dbr}, respectively.
Substituting the zeroth-order propagators into Eq.~\eqref{eq:VIA_I_diamond0_alt}, which is equivalent to Eq.~\eqref{eq:VIA_I_diamond0}, we obtain
\begin{align}
  \left. \mathcal{I}^{\lambda} \right|_{\diamond^0} &= -\alpha_-G^{\lambda,[0]} +g^\lambda G^{a,[0]} +G^{\lambda,[0]}\alpha_+ -G^{r,[0]}g^\lambda \notag \\*
  &= 0.
  \label{eq:VIA_cancellation_identity}
\end{align}
In the second equality, we have used
\begin{align}
  \alpha_- G^{\lambda,[0]} - g^\lambda G^{a,[0]} = 0, \qquad
  G^{\lambda,[0]} \alpha_+ - G^{r,[0]} g^\lambda = 0,
  \label{eq:VIA_left_right}
\end{align}
which are precisely the zeroth-order forms of Eq.~\eqref{eq:Wightman2} and its conjugate equation, respectively.
Consequently, both the bosonic and fermionic source terms defined in Eqs.~\eqref{eq:VIA_boson_source} and \eqref{eq:VIA_fermion_source} vanish.

This result corresponds to the VIA cancellation found in ref.~\cite{Postma:2022dbr}.
The VIA cancellation therefore follows from the identity in Eq.~\eqref{eq:VIA_cancellation_identity}.
This identity is a linear combination of Eq.~\eqref{eq:Wightman2} and its conjugate equation at zeroth order in the derivative expansion, where the term involving $\mathcal{D}_K$ is absent.
Accordingly, the source terms in Eqs.~\eqref{eq:VIA_boson_source} and \eqref{eq:VIA_fermion_source} vanish when evaluated with the zeroth-order solution.
Consequently, retaining the current-divergence term involving $\mathcal{D}_K$ on the left-hand side in Eqs.~\eqref{eq:VIA_bosonic_currentdiv} and \eqref{eq:VIA_fermionic_currentdiv}, while evaluating the right-hand side only at zeroth order, does not provide an additional equation that determines the current divergence; the VIA cancellation follows from the zeroth-order equations in Eq.~\eqref{eq:VIA_left_right} that have been used to determine the solution.
Moreover, the zeroth-order solution does not contain the effects of spacetime derivatives of the mass and therefore does not address the effects of the wall gradients on the current divergence.
This result alone does not demonstrate the absence of wall-gradient contributions to the VIA source.
Therefore, the criticism of the VIA in ref.~\cite{Postma:2022dbr} is conceptually different from the issue identified in ref.~\cite{Kainulainen:2021oqs}, which concerns the treatment of the spacetime-dependent mass in the self-energy.

In our analysis presented in the main text, we obtain the solutions that include the wall-gradient effects by carrying out the derivative expansion up to first order in the presence of self-energy corrections (see section~\ref{sec:solution}).
Under the conditions for deriving the Boltzmann equation discussed in section~\ref{sec:approx}, the effects of spacetime derivatives of the mass are consistently incorporated into the delta-function structure of the solutions, thereby modifying the on-shell condition for the quasiparticle.
As a result, our approach reproduces the results of the semiclassical approach without treating the spacetime-dependent mass through a nonlocal insertion in the self-energy, as is done in the conventional VIA.

\section{Lorentz boost \label{sec:boost}}

In this appendix, we summarize the Lorentz boost from the wall frame to the boosted frame.
The boost is defined by
\begin{align}
  \tilde{k}^\mu = \Lambda_\parallel{}^\mu{}_\nu k^\nu . 
\end{align}
The nonzero components of the boost matrix are
\begin{align}
  \Lambda_\parallel{}^0{}_0 &=
  \gamma_\parallel,
  ~~~~
  \Lambda_\parallel{}^0{}_a =
  \Lambda_\parallel{}^a{}_0 = -\gamma_\parallel \beta^a, ~~~~ 
  \Lambda_\parallel{}^a{}_b =
  \delta^a{}_b + (\gamma_\parallel-1) \frac{\beta^a\beta^b}{|\bm{\beta}|^2},
  ~~~~
  \Lambda_\parallel{}^3{}_3 = 1,
  \label{eq:boost_parallel_matrix}
\end{align}
where $\beta^a = k^a/k^0$ with $a,b=1,2$.
With this choice, one obtains
\begin{align}
  \tilde{\bm{k}}_\parallel = 0,
  ~~~~
  \tilde{k}_z = k_z,
  ~~~~
  \tilde{k}^0 = \gamma_\parallel^{-1}k^0.
  \label{eq:boost_parallel_result}
\end{align}
The same boost is applied to the plasma four-velocity, $\tilde{u}^\mu = \Lambda_\parallel{}^\mu{}_\nu u^\nu$.
The spinor decomposition introduced in section~\ref{sec:spindecompose} is defined in the boosted frame.

\section{Hermiticity and discrete symmetries \label{sec:hermiticity}}

In this appendix, we summarize the hermiticity relations and the transformation laws under the discrete symmetries for bosons and fermions.

\subsection{Hermiticity}

The hermiticity relations of the Wightman functions are
\begin{align}
  &(i \Delta^{\lambda}_{ij} (u,v))^\dagger = i \Delta^{\lambda}_{ji} (v,u), \notag \\
  &\big( (\gamma^0)_{\alpha \delta} i S^{\lambda}_{\delta \beta, ij} (u,v) \big)^\dagger = (\gamma^0)_{\beta \delta} i S^{\lambda}_{\delta \alpha, ji} (v,u),
\end{align}
where we have explicitly shown the flavor indices and the spinor indices.
For fermions, this can equivalently be written as
\begin{align}
  \big( i S^{\lambda}_{\alpha \delta, ij} (u,v) (\gamma^0)_{\delta \beta} \big)^\dagger = i S^{\lambda}_{\beta \delta, ji} (v,u)  (\gamma^0)_{\delta \alpha}.
\end{align}
In the Wigner representation, these hermiticity relations become
\begin{align}
  &(i \Delta^{\lambda}_{ij} (k,x))^\dagger = i \Delta^{\lambda}_{ji} (k,x), \notag \\
  &\big(  (\gamma^0)_{\alpha \delta}  i S^{\lambda}_{\delta \beta, ij} (k,x) \big)^\dagger = (\gamma^0)_{\beta \delta} i S^{\lambda}_{\delta \alpha ,ji} (k,x).
\end{align}

\subsection{C and CP transformations}

We define the C and P transformations in the flavor basis.
The C transformation for bosonic and fermionic fields is defined by
\begin{align}
  \phi \xrightarrow{\mathrm{C}} \phi^\dagger, \qquad \psi \xrightarrow{\mathrm{C}} C \overline{\psi}^\intercal,
\end{align}
where $C = i \gamma^0 \gamma^2$. 
The P transformation is defined by
\begin{align}
  \phi (u) \xrightarrow{\mathrm{P}} \phi (u_P), \qquad  \psi (u) \xrightarrow{\mathrm{P}} \gamma^0 \psi (u_P),
\end{align}
where $u_P = (u^0, -\bm{u})$. 
We then obtain
\begin{align}
  i \Delta^\lambda_{ij} (u,v) &\xrightarrow{\mathrm{C}} i \Delta_{ji}^{\overline{\lambda}} (v,u), \notag \\
  i \Delta^\lambda_{ij} (u,v) &\xrightarrow{\mathrm{CP}} i \Delta_{ji}^{\overline{\lambda}} (v_P,u_P), \notag \\
  i S^\lambda_{\alpha \beta, ij} (u,v) &\xrightarrow{\mathrm{C}} - C_{\alpha \delta} iS_{\rho \delta, ji}^{\overline{\lambda}} (v,u) C_{\rho \beta}, \notag \\
  i S^\lambda_{\alpha \beta, ij} (u,v) &\xrightarrow{\mathrm{CP}} - (C\gamma^0)_{\alpha \delta} iS_{\rho \delta, ji}^{\overline{\lambda}} (v_P,u_P) (\gamma^0 C)_{\rho \beta} 
\end{align} 
where $\overline{\lambda}$ denotes the opposite Wightman component, namely $\overline{\lambda}=>$ for $\lambda=<$ and $\overline{\lambda}=<$ for $\lambda=>$.
In Wigner space, we have 
\begin{align}
  i \Delta^\lambda_{ ij} (k,x) &\xrightarrow{\mathrm{C}} ~ i\Delta_{ ji}^{\overline{\lambda}} (-k,x), \notag \\
  i \Delta^\lambda_{ ij} (k,x) &\xrightarrow{\mathrm{CP}} i\Delta_{ji}^{\overline{\lambda}} (-k_P,x_P), \notag \\
  i S^\lambda_{\alpha \beta, ij} (k,x) &\xrightarrow{\mathrm{C}} - C_{\alpha \delta} iS_{\rho \delta, ji}^{\overline{\lambda}} (-k,x) C_{\rho \beta}, \notag \\
  i S^\lambda_{\alpha \beta, ij} (k,x) &\xrightarrow{\mathrm{CP}} - (C\gamma^0)_{\alpha \delta} iS_{\rho \delta, ji}^{\overline{\lambda}} (-k_P,x_P) (\gamma^0 C)_{\rho \beta} .
\end{align}
The corresponding self-energies obey the same transformation relations.

When the interactions encoded in the self-energies are CP invariant, the self-energy satisfies
\begin{align}
  \Pi^\lambda (k) = [\Pi^{\overline{\lambda}} (-k_P)]^\intercal, ~~~~~~ \Sigma^\lambda (k) = -\gamma^2 [\Sigma^{\overline{\lambda}} (-k_P)]^\intercal \gamma^2,
\end{align}
where the transpose is taken over flavor indices for $\Pi^\lambda$ and over both flavor and spinor indices for $\Sigma^\lambda$.
Using the definitions of the Hermitian and anti-Hermitian parts, we obtain
\begin{align}
  &\Pi_{\mathcal{A}}(- k_P) = -[\Pi_{\mathcal{A}} (k)]^\intercal, \qquad \qquad \Pi_{\mathcal{H}}(- k_P) = [\Pi_{\mathcal{H}} (k)]^\intercal, \notag \\
  &\Sigma_{\mathcal{A}}(- k_P) = \gamma^2 [\Sigma_{\mathcal{A}} (k)]^\intercal \gamma^2, \qquad \quad \Sigma_{\mathcal{H}}(- k_P) = - \gamma^2 [\Sigma_{\mathcal{H}} (k)]^\intercal \gamma^2.
\end{align}

\bibliographystyle{utphys31mod}

\bibliography{biblio}

\providecommand{\href}[2]{#2}\begingroup\begin{thebibliography}{10}

\bibitem{Kuzmin:1985mm}
V.~A.~Kuzmin, V.~A.~Rubakov, and M.~E.~Shaposhnikov, ``{On the Anomalous Electroweak Baryon Number Nonconservation in the Early Universe},'' \href{https://doi.org/10.1016/0370-2693(85)91028-7}{Phys.\  Lett.\  B {\bfseries 155} (1985) 36}.

\bibitem{Cohen:1994ss}
A.~G.~Cohen, D.~B.~Kaplan, and A.~E.~Nelson, ``{Diffusion enhances spontaneous electroweak baryogenesis},'' \href{https://doi.org/10.1016/0370-2693(94)00935-X}{Phys.\  Lett.\  B {\bfseries 336} (1994) 41--47} {\ttfamily [\href{https://arxiv.org/abs/hep-ph/9406345}{hep-ph/9406345}]}.

\bibitem{Joyce:1994fu}
M.~Joyce, T.~Prokopec, and N.~Turok, ``{Electroweak baryogenesis from a classical force},'' \href{https://doi.org/10.1103/PhysRevLett.75.1695}{Phys.\  Rev.\  Lett.\  {\bfseries 75} (1995) 1695--1698} {\ttfamily [\href{https://arxiv.org/abs/hep-ph/9408339}{hep-ph/9408339}]}. [Erratum: Phys.Rev.Lett. 75, 3375 (1995)].

\bibitem{Joyce:1994zt}
M.~Joyce, T.~Prokopec, and N.~Turok, ``{Nonlocal electroweak baryogenesis. Part 2: The Classical regime},'' \href{https://doi.org/10.1103/PhysRevD.53.2958}{Phys.\  Rev.\  D {\bfseries 53} (1996) 2958--2980} {\ttfamily [\href{https://arxiv.org/abs/hep-ph/9410282}{hep-ph/9410282}]}.

\bibitem{Huet:1995sh}
P.~Huet and A.~E.~Nelson, ``{Electroweak baryogenesis in supersymmetric models},'' \href{https://doi.org/10.1103/PhysRevD.53.4578}{Phys.\  Rev.\  D {\bfseries 53} (1996) 4578--4597} {\ttfamily [\href{https://arxiv.org/abs/hep-ph/9506477}{hep-ph/9506477}]}.

\bibitem{Cline:1997vk}
J.~M.~Cline, M.~Joyce, and K.~Kainulainen, ``{Supersymmetric electroweak baryogenesis in the WKB approximation},'' \href{https://doi.org/10.1016/S0370-2693(97)01361-0}{Phys.\  Lett.\  B {\bfseries 417} (1998) 79--86} {\ttfamily [\href{https://arxiv.org/abs/hep-ph/9708393}{hep-ph/9708393}]}. [Erratum: Phys.Lett.B 448, 321--321 (1999)].

\bibitem{Cline:2000nw}
J.~M.~Cline, M.~Joyce, and K.~Kainulainen, ``{Supersymmetric electroweak baryogenesis},'' \href{https://doi.org/10.1088/1126-6708/2000/07/018}{JHEP {\bfseries 07} (2000) 018} {\ttfamily [\href{https://arxiv.org/abs/hep-ph/0006119}{hep-ph/0006119}]}.

\bibitem{Kainulainen:2001cn}
K.~Kainulainen, T.~Prokopec, M.~G.~Schmidt, and S.~Weinstock, ``{First principle derivation of semiclassical force for electroweak baryogenesis},'' \href{https://doi.org/10.1088/1126-6708/2001/06/031}{JHEP {\bfseries 06} (2001) 031} {\ttfamily [\href{https://arxiv.org/abs/hep-ph/0105295}{hep-ph/0105295}]}.

\bibitem{Kainulainen:2002th}
K.~Kainulainen, T.~Prokopec, M.~G.~Schmidt, and S.~Weinstock, ``{Semiclassical force for electroweak baryogenesis: Three-dimensional derivation},'' \href{https://doi.org/10.1103/PhysRevD.66.043502}{Phys.\  Rev.\  D {\bfseries 66} (2002) 043502} {\ttfamily [\href{https://arxiv.org/abs/hep-ph/0202177}{hep-ph/0202177}]}.

\bibitem{Prokopec:2003pj}
T.~Prokopec, M.~G.~Schmidt, and S.~Weinstock, ``{Transport equations for chiral fermions to order $\hbar$ and electroweak baryogenesis. Part 1},'' \href{https://doi.org/10.1016/j.aop.2004.06.002}{Annals Phys.\  {\bfseries 314} (2004) 208--265} {\ttfamily [\href{https://arxiv.org/abs/hep-ph/0312110}{hep-ph/0312110}]}.

\bibitem{Prokopec:2004ic}
T.~Prokopec, M.~G.~Schmidt, and S.~Weinstock, ``{Transport equations for chiral fermions to order $\hbar$ and electroweak baryogenesis. Part II},'' \href{https://doi.org/10.1016/j.aop.2004.06.001}{Annals Phys.\  {\bfseries 314} (2004) 267--320} {\ttfamily [\href{https://arxiv.org/abs/hep-ph/0406140}{hep-ph/0406140}]}.

\bibitem{Fromme:2006wx}
L.~Fromme and S.~J.~Huber, ``{Top transport in electroweak baryogenesis},'' \href{https://doi.org/10.1088/1126-6708/2007/03/049}{JHEP {\bfseries 03} (2007) 049} {\ttfamily [\href{https://arxiv.org/abs/hep-ph/0604159}{hep-ph/0604159}]}.

\bibitem{Cline:2020jre}
J.~M.~Cline and K.~Kainulainen, ``{Electroweak baryogenesis at high bubble wall velocities},'' \href{https://doi.org/10.1103/PhysRevD.101.063525}{Phys.\  Rev.\  D {\bfseries 101} (2020) 063525} {\ttfamily [\href{https://arxiv.org/abs/2001.00568}{arXiv:2001.00568}]}.

\bibitem{Kainulainen:2021oqs}
K.~Kainulainen, ``{CP-violating transport theory for electroweak baryogenesis with thermal corrections},'' \href{https://doi.org/10.1088/1475-7516/2021/11/042}{JCAP {\bfseries 11} (2021) 042} {\ttfamily [\href{https://arxiv.org/abs/2108.08336}{arXiv:2108.08336}]}.

\bibitem{Kainulainen:2024qpm}
K.~Kainulainen and N.~Venkatesan, ``{Systematic moment expansion for electroweak baryogenesis},'' \href{https://doi.org/10.1088/1475-7516/2024/08/058}{JCAP {\bfseries 08} (2024) 058} {\ttfamily [\href{https://arxiv.org/abs/2407.13639}{arXiv:2407.13639}]}.

\bibitem{Schwinger:1960qe}
J.~S.~Schwinger, ``{Brownian motion of a quantum oscillator},'' \href{https://doi.org/10.1063/1.1703727}{J.\  Math.\  Phys.\  {\bfseries 2} (1961) 407--432}.

\bibitem{Keldysh:1964ud}
L.~V.~Keldysh, ``{Diagram Technique for Nonequilibrium Processes},'' \href{https://doi.org/10.1142/9789811279461_0007}{Sov.\  Phys.\  JETP {\bfseries 20} (1965) 1018--1026}.

\bibitem{Kadanoff:1962}
L.~P.~Kadanoff and G.~Baym, {\em Quantum Statistical Mechanics: Green's Function Methods in Equilibrium and Nonequilibrium Problems}.
\newblock W. A. Benjamin, New York, 1962.

\bibitem{Konstandin:2004gy}
T.~Konstandin, T.~Prokopec, and M.~G.~Schmidt, ``{Kinetic description of fermion flavor mixing and CP-violating sources for baryogenesis},'' \href{https://doi.org/10.1016/j.nuclphysb.2005.03.013}{Nucl.\  Phys.\  B {\bfseries 716} (2005) 373--400} {\ttfamily [\href{https://arxiv.org/abs/hep-ph/0410135}{hep-ph/0410135}]}.

\bibitem{Konstandin:2005cd}
T.~Konstandin, T.~Prokopec, M.~G.~Schmidt, and M.~Seco, ``{MSSM electroweak baryogenesis and flavor mixing in transport equations},'' \href{https://doi.org/10.1016/j.nuclphysb.2005.11.028}{Nucl.\  Phys.\  B {\bfseries 738} (2006) 1--22} {\ttfamily [\href{https://arxiv.org/abs/hep-ph/0505103}{hep-ph/0505103}]}.

\bibitem{Herranen:2008hi}
M.~Herranen, K.~Kainulainen, and P.~M.~Rahkila, ``{Towards a kinetic theory for fermions with quantum coherence},'' \href{https://doi.org/10.1016/j.nuclphysb.2008.09.032}{Nucl.\  Phys.\  B {\bfseries 810} (2009) 389--426} {\ttfamily [\href{https://arxiv.org/abs/0807.1415}{arXiv:0807.1415}]}.

\bibitem{Cirigliano:2009yt}
V.~Cirigliano, C.~Lee, M.~J.~Ramsey-Musolf, and S.~Tulin, ``{Flavored Quantum Boltzmann Equations},'' \href{https://doi.org/10.1103/PhysRevD.81.103503}{Phys.\  Rev.\  D {\bfseries 81} (2010) 103503} {\ttfamily [\href{https://arxiv.org/abs/0912.3523}{arXiv:0912.3523}]}.

\bibitem{Cirigliano:2011di}
V.~Cirigliano, C.~Lee, and S.~Tulin, ``{Resonant Flavor Oscillations in Electroweak Baryogenesis},'' \href{https://doi.org/10.1103/PhysRevD.84.056006}{Phys.\  Rev.\  D {\bfseries 84} (2011) 056006} {\ttfamily [\href{https://arxiv.org/abs/1106.0747}{arXiv:1106.0747}]}.

\bibitem{Riotto:1995hh}
A.~Riotto, ``{Towards a nonequilibrium quantum field theory approach to electroweak baryogenesis},'' \href{https://doi.org/10.1103/PhysRevD.53.5834}{Phys.\  Rev.\  D {\bfseries 53} (1996) 5834--5841} {\ttfamily [\href{https://arxiv.org/abs/hep-ph/9510271}{hep-ph/9510271}]}.

\bibitem{Riotto:1997vy}
A.~Riotto, ``{Supersymmetric electroweak baryogenesis, nonequilibrium field theory and quantum Boltzmann equations},'' \href{https://doi.org/10.1016/S0550-3213(98)00159-X}{Nucl.\  Phys.\  B {\bfseries 518} (1998) 339--360} {\ttfamily [\href{https://arxiv.org/abs/hep-ph/9712221}{hep-ph/9712221}]}.

\bibitem{Riotto:1998zb}
A.~Riotto, ``{The More relaxed supersymmetric electroweak baryogenesis},'' \href{https://doi.org/10.1103/PhysRevD.58.095009}{Phys.\  Rev.\  D {\bfseries 58} (1998) 095009} {\ttfamily [\href{https://arxiv.org/abs/hep-ph/9803357}{hep-ph/9803357}]}.

\bibitem{Postma:2022dbr}
M.~Postma, J.~van~de Vis, and G.~White, ``{Resummation and cancellation of the VIA source in electroweak baryogenesis},'' \href{https://doi.org/10.1007/JHEP12(2022)121}{JHEP {\bfseries 12} (2022) 121} {\ttfamily [\href{https://arxiv.org/abs/2206.01120}{arXiv:2206.01120}]}.

\bibitem{Laurent:2020gpg}
B.~Laurent and J.~M.~Cline, ``{Fluid equations for fast-moving electroweak bubble walls},'' \href{https://doi.org/10.1103/PhysRevD.102.063516}{Phys.\  Rev.\  D {\bfseries 102} (2020) 063516} {\ttfamily [\href{https://arxiv.org/abs/2007.10935}{arXiv:2007.10935}]}.

\bibitem{Dorsch:2021ubz}
G.~C.~Dorsch, S.~J.~Huber, and T.~Konstandin, ``{On the wall velocity dependence of electroweak baryogenesis},'' \href{https://doi.org/10.1088/1475-7516/2021/08/020}{JCAP {\bfseries 08} (2021) 020} {\ttfamily [\href{https://arxiv.org/abs/2106.06547}{arXiv:2106.06547}]}.

\bibitem{Laurent:2022jrs}
B.~Laurent and J.~M.~Cline, ``{First principles determination of bubble wall velocity},'' \href{https://doi.org/10.1103/PhysRevD.106.023501}{Phys.\  Rev.\  D {\bfseries 106} (2022) 023501} {\ttfamily [\href{https://arxiv.org/abs/2204.13120}{arXiv:2204.13120}]}.

\bibitem{Garbrecht:2011xw}
B.~Garbrecht and M.~Garny, ``{Finite Width in out-of-Equilibrium Propagators and Kinetic Theory},'' \href{https://doi.org/10.1016/j.aop.2011.10.005}{Annals Phys.\  {\bfseries 327} (2012) 914--934} {\ttfamily [\href{https://arxiv.org/abs/1108.3688}{arXiv:1108.3688}]}.

\bibitem{Joyce:1999fw}
M.~Joyce, K.~Kainulainen, and T.~Prokopec, ``{The Semiclassical propagator in field theory},'' \href{https://doi.org/10.1016/S0370-2693(99)01169-7}{Phys.\  Lett.\  B {\bfseries 468} (1999) 128--133} {\ttfamily [\href{https://arxiv.org/abs/hep-ph/9906411}{hep-ph/9906411}]}.

\bibitem{Joyce:1999uf}
M.~Joyce, K.~Kainulainen, and T.~Prokopec, ``{Quantum transport equations for a scalar field},'' \href{https://doi.org/10.1016/S0370-2693(00)00041-1}{Phys.\  Lett.\  B {\bfseries 474} (2000) 402--410} {\ttfamily [\href{https://arxiv.org/abs/hep-ph/9910535}{hep-ph/9910535}]}.

\bibitem{Cline:2021dkf}
J.~M.~Cline and B.~Laurent, ``{Electroweak baryogenesis from light fermion sources: A critical study},'' \href{https://doi.org/10.1103/PhysRevD.104.083507}{Phys.\  Rev.\  D {\bfseries 104} (2021) 083507} {\ttfamily [\href{https://arxiv.org/abs/2108.04249}{arXiv:2108.04249}]}.

\bibitem{Basler:2021kgq}
P.~Basler, L.~Biermann, M.~M{\"u}hlleitner, and J.~M{\"u}ller, ``{Electroweak baryogenesis in the CP-violating two-Higgs doublet model},'' \href{https://doi.org/10.1140/epjc/s10052-023-11192-9}{Eur.\  Phys.\  J.\  C {\bfseries 83} (2023) 57} {\ttfamily [\href{https://arxiv.org/abs/2108.03580}{arXiv:2108.03580}]}.

\bibitem{Postma:2019scv}
M.~Postma and J.~van~de Vis, ``{Source terms for electroweak baryogenesis in the vev-insertion approximation beyond leading order},'' \href{https://doi.org/10.1007/JHEP02(2020)090}{JHEP {\bfseries 02} (2020) 090} {\ttfamily [\href{https://arxiv.org/abs/1910.11794}{arXiv:1910.11794}]}.

\bibitem{Postma:2021zux}
M.~Postma, ``{A different perspective on the vev insertion approximation for electroweak baryogenesis},'' \href{https://doi.org/10.1007/JHEP09(2021)055}{JHEP {\bfseries 09} (2021) 055} {\ttfamily [\href{https://arxiv.org/abs/2107.05971}{arXiv:2107.05971}]}.

\bibitem{vandeVis:2025efm}
J.~van~de Vis, J.~de~Vries, and M.~Postma, ``{Bubble Trouble: a Review on Electroweak Baryogenesis}.'' {\ttfamily \href{https://arxiv.org/abs/2508.09989}{arXiv:2508.09989}}.

\end{thebibliography}\endgroup

\end{document}